\documentclass[11pt,oneside,letterpaper]{article}
\usepackage{amssymb}
\usepackage{amsmath,bm}
\usepackage[dvips]{graphicx}
\usepackage{setspace}
\usepackage{fancyhdr}
\usepackage[dvipsnames]{xcolor}
\usepackage{ifpdf}
\usepackage{graphicx}
\usepackage{rotating}
\usepackage{comment}
\usepackage[margin=2.5cm]{geometry}
\usepackage[colorlinks=true, linkcolor  = blue,urlcolor=blue,citecolor=violet]{hyperref}
\usepackage{soul}
\usepackage{relsize}
\usepackage{tikz}
\usepackage{float}
\usepackage{empheq}
\usepackage{bbold}
\usepackage{cite}
\usepackage{etoolbox}
\patchcmd{\thebibliography}{\section*{\refname}}{}{}{}

\newcommand{\bi}{\begin{itemize}}
\newcommand{\ei}{\end{itemize}}
\newcommand{\bea}{\begin{eqnarray}}
\newcommand{\eea}{\end{eqnarray}}
\newcommand{\be}{\begin{equation}}
\newcommand{\ee}{\end{equation}}

\newcommand{\dd}{\mathrm{d}}

\newcommand{\PP}{\mathcal{P}}
\numberwithin{equation}{section}
\allowdisplaybreaks  

\begin{document}

\begin{center}

~
\vskip4mm
{{\huge {The semiclassical gravitational path integral\\  and random matrices}
\quad 
\\  \vskip5mm
\Large {(Toward a microscopic picture of a dS$_2$ universe)}
 
  }}
\vskip5mm

\vskip2mm

\vskip10mm


Dionysios Anninos$^{\includegraphics[scale=.008]{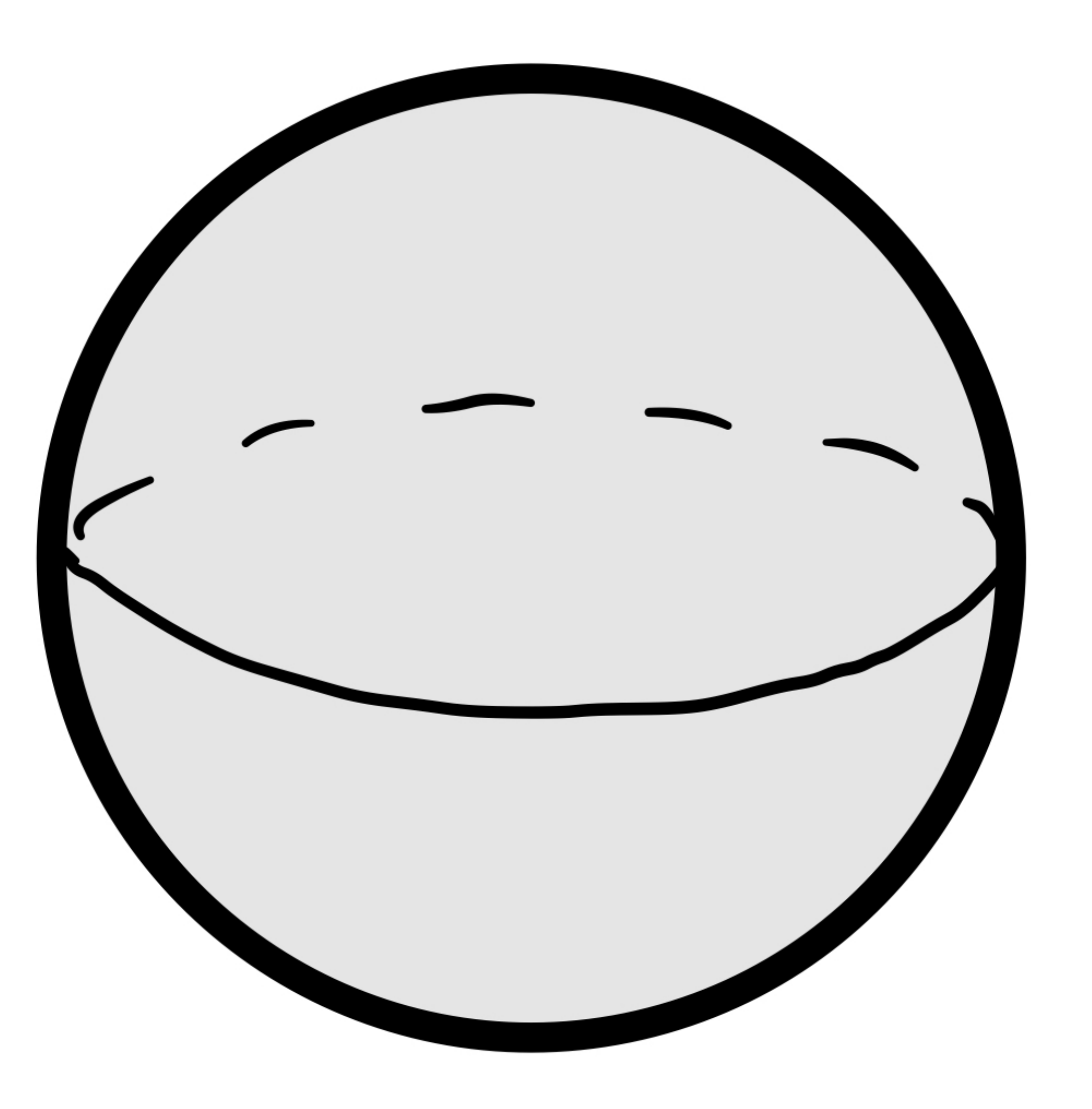}}$ $\&$~Beatrix M\"uhlmann$^{\includegraphics[scale=.008]{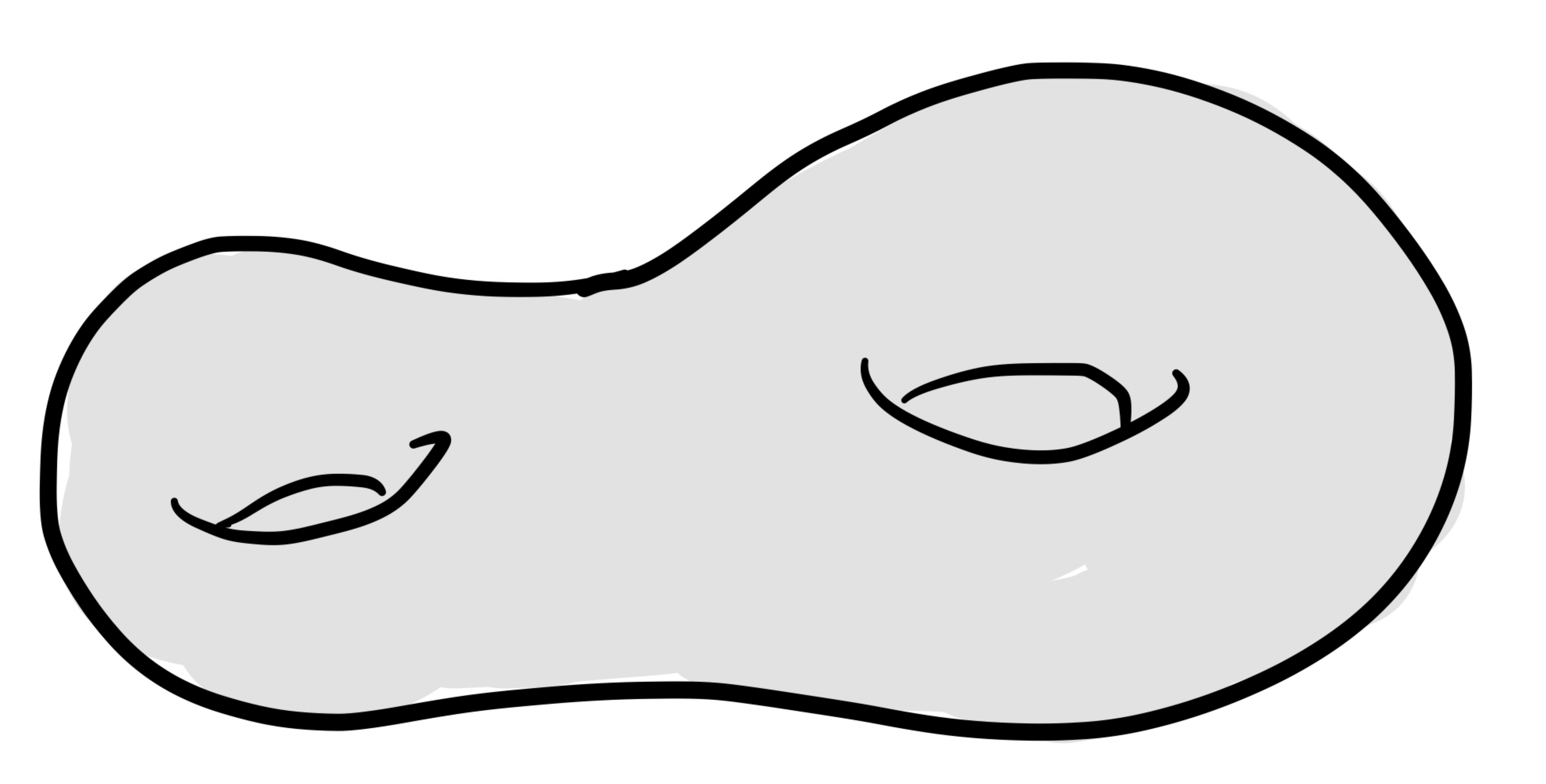}}$ \\ 

\end{center}
\vskip4mm
\begin{center}
{
\footnotesize
{$^{\includegraphics[scale=.008]{test}}$Department of Mathematics, King's College London, Strand, London WC2R 2LS, UK \newline\newline
$^{\includegraphics[scale=.008]{test2}}$Department of Physics, McGill University, Montreal QC H3A 2T8, Canada\\
}}
\end{center}
\begin{center}
{\textsf{\footnotesize{
dionysios.anninos@kcl.ac.uk, beatrix.muehlmann@mcgill.ca}} } 
\end{center}

\vspace*{0.5cm}

\vspace*{1.5cm}
\begin{abstract}
\noindent
\\ \\ 
We study the genus expansion on compact Riemann surfaces of the gravitational path integral $\mathcal{Z}^{(m)}_{\text{grav}}$ in two spacetime dimensions with cosmological constant $\Lambda>0$ coupled to one of the non-unitary minimal models $\mathcal{M}_{2m-1,2}$. In the semiclassical limit, corresponding to large $m$, $\mathcal{Z}^{(m)}_{\text{grav}}$ admits a Euclidean saddle for genus $h\geq 2$.
Upon fixing the area of the metric, the path integral admits a round two-sphere saddle for $h=0$. We show that the OPE coefficients for the minimal weight operators of $\mathcal{M}_{2m-1,2}$ grow exponentially in $m$ at large $m$. Employing the sewing formula, we use these OPE coefficients  to obtain the large $m$ limit of the partition function of $\mathcal{M}_{2m-1,2}$ for genus $h\ge 2$. Combining these results we arrive at a semiclassical expression for $\mathcal{Z}^{(m)}_{\text{grav}}$.
Conjecturally, $\mathcal{Z}^{(m)}_{\text{grav}}$ admits a completion in terms of an integral over large random Hermitian matrices, known as a multicritical matrix integral. This matrix integral is built from an even polynomial potential of order $2m$. We obtain explicit expressions for the large $m$ genus expansion of multicritical matrix integrals in the double scaling limit. We compute invariant quantities involving contributions at different genera, both from a matrix as well as a gravity perspective, and establish a link between the two pictures. Inspired by the proposal of Gibbons and Hawking relating the de Sitter entropy to a gravitational path integral, our setup paves a possible path toward a microscopic picture of a  two-dimensional de Sitter universe. 




\end{abstract}

\newpage
\setcounter{page}{1}
\pagenumbering{arabic}

\tableofcontents

\setcounter{page}{1}
\pagenumbering{arabic}

\onehalfspacing

\section{Introduction}

In  \cite{Gibbons:1976ue,Gibbons:1977mu} Gibbons and Hawking postulate that the entropy $S_{\text{dS}}$ of the cosmological event horizon of a de Sitter universe is macroscopically defined by the Euclidean path integral
\begin{equation}\label{GH}
e^{S_{\text{dS}}}  =  \sum_{\mathcal{M}} \int  [\mathcal{D} g ] e^{-S_E[\Lambda,g_{ij};\mathcal{M}] }  Z_{\text{matter}}[g_{ij};\mathcal{M}]~.
\end{equation}
Here $S_E[\Lambda,g_{ij};\mathcal{M}]$ denotes the Euclidean gravitational action in $(d+1)$ spacetime dimensions with $d\geq 2$ and positive cosmological constant $\Lambda>0$. The sum is over compact manifolds $\mathcal{M}$. The dominant saddle of (\ref{GH}) is the round $(d+1)$-sphere, i.e. Euclidean $(d+1)$-dimensional de Sitter space. Matter fields contribute through the matter partition function $Z_{\text{matter}}[g_{ij};\mathcal{M}]$. Expression (\ref{GH}) can be viewed as a cosmological version of Euclidean black hole thermodynamics. The absence of an energy term on the left hand side of (\ref{GH}) is due to the compactness of the Cauchy surfaces in a de Sitter spacetime. A recent and detailed exploration of (\ref{GH}) is found in  \cite{longsphere,timelike,Law:2020cpj,David:2021wrw,Anninos:2021ihe,Hikida:2021ese}. 
Foregoing (at least for now) phenomenologically motivated models, and guided by the necessity for a toy model of a quantum de Sitter spacetime in any dimension, this paper explores the details of the Gibbons-Hawking path integral (\ref{GH}) extended to the case of two spacetime dimensions \cite{musings,penguins,beatrix}.

An important motivation for focusing on two dimensions is the conjectural completion of (\ref{GH}) in terms of an integral over large random Hermitian matrices.
The class of matrix integrals $\mathcal{M}_N^{(m)}(\boldsymbol{\alpha})$ we focus on are known as multicritical matrix integrals. 
These are matrix integrals built out of a Hermitian $N\times N$ matrix organised in an even polynomial  potential of order $2m$ with $(m-1)$ real valued couplings $\boldsymbol{\alpha} \in \mathbb{R}^{m-1}$. In the limit where $N$ tends to infinity, while simultaneously tuning the couplings to the multicritical point $\boldsymbol{\alpha}_c^{(m)}$ (the double scaling limit) \cite{Douglas:1989ve,Gross:1989vs,Brezin:1990rb}, multicritical matrix integrals are conjectured \cite{Staudacher:1989fy,Kazakov:1989bc,Moore:1991ir} to be dual to two-dimensional quantum gravity coupled to the  $\mathcal{M}_{2m-1,2}$ series of non-unitary minimal models\footnote{Other contexts where a continuum field theoretic path integral localises to a matrix integral include supersymmetric localisation \cite{Pestun:2016zxk}, Chern-Simons theory \cite{Marino:2002fk,Tierz:2002jj}, and the thermal path integral of certain fermionic  quantum mechanical matrix models \cite{Anninos:2016klf}. In all these cases, the matrix integral is associated to theories with an underlying Hilbert space.}
\begin{equation}\label{M_grav}
-\lim_{\text{d.sc.}} \log\mathcal{M}_N^{(m)}(\boldsymbol{\alpha}) = \mathcal{N}\times \mathcal{Z}_{\text{grav}}^{(m)}[\Lambda,\vartheta]~.
\end{equation}
Here $\mathcal{N}$ denotes a normalisation constant and the gravitational path integral (\ref{GH}) in two dimensions is given by
\begin{equation}\label{Zgrav_intro}
\mathcal{Z}^{(m)}_{\text{grav}}[\Lambda,\vartheta] \equiv \sum_{h=0}^\infty e^{\vartheta \chi_h} \int [\mathcal{D} g] e^{-\Lambda \int_{\Sigma_h} \dd^2 x\sqrt{g}} \, Z_{\text{CFT}}^{(m)}[g_{ij} ;\Sigma_h]~,
\end{equation}
where $\chi_h= 2-2h$ is the Euler characteristic of a compact Riemann surface $\Sigma_h$ of genus $h$, while $\Lambda$ couples the identity operator  of the minimal model to gravity. The remaining $(m-2)$ Virasoro primaries of $\mathcal{M}_{2m-1,2}$ are switched off in (\ref{Zgrav_intro}). There exists a concrete dictionary \cite{penguins} between the $(m-1)$ primaries of $\mathcal{M}_{2m-1,2}$ and the $(m-1)$ couplings $\boldsymbol{\alpha}$ of $\mathcal{M}_N^{(m)}(\boldsymbol{\alpha})$.

In general, the  fluctuations of the metric field in the gravitational path integral (\ref{Zgrav_intro}) are unsuppressed. On a genus zero surface we can tame these fluctuations by fixing the area of the physical metric. In the large $m$ limit, corresponding to a large and negative matter central charge, the gravitational path integral exhibits a semiclassical round two-sphere saddle \cite{Zamolodvarphikov:1982vx}. Higher topologies are suppressed for large $\vartheta$. For $h\geq 2$ the gravitational path integral (\ref{Zgrav_intro}) admits a semiclassical large $m$ saddle without any fixed area constraint \cite{Takhtajan:2005md}.

Granting the conjecture of \cite{Staudacher:1989fy,Kazakov:1989bc,Moore:1991ir}, the matrix integral $\mathcal{M}^{(m)}_N(\boldsymbol{\alpha})$ yields a microscopic realisation of the Gibbons-Hawking path integral in two spacetime dimensions. Interestingly, from the perspective of the planar expansion of the matrix integral the inclusion of non-spherical compact topologies in $\mathcal{Z}^{(m)}_{\text{grav}}[\Lambda,\vartheta] $ is not optional but necessary. ({In $(d+1)$-dimensions with $d\ge 3$, there exists circumstantial evidence  \cite{Ginsparg:1982rs}  that at least the $S^2 \times S^{d-1}$ Nariai instanton should  be included in the Gibbons-Hawking path integral.})
 Further to this, non-perturbative effects \cite{Polchinski:1994fq,Ginsparg:1991ws,Eynard:1992sg,Shenker:1990uf} necessitate contributions of order $e^{\vartheta \chi_{h,b} }$ with $\chi_{h,b} = 2-2h-b$, which can be achieved through the inclusion of manifolds $\Sigma_{h,b}$ with $b$ boundaries. Perhaps such effects modify the original boundary conditions of the Gibbons-Hawking path integral (\ref{GH}).  

\subsection*{Outline $\&$ results}

In the following sections we explore the relation (\ref{M_grav}) in the semiclassical, large $m$, limit for genus $h\geq 0$. We collect technical details in the various appendices. 
\newline\newline
In section \ref{matrix} we review the main properties of large $m$ multicritical matrix integrals $\mathcal{M}_N^{(m)}(\boldsymbol{\alpha})$ and study the large $m$ behaviour of the genus expansion of $ \log\mathcal{M}_N^{(m)}(\boldsymbol{\alpha})$ as we approach the multicritical point $\boldsymbol{\alpha}_c^{(m)}$ from different directions.  In particular we identify the path corresponding to the cosmological constant $\Lambda$, which couples the identity operator of $\mathcal{M}_{2m-1,2}$ to gravity. The resulting expression for the genus $h$ contribution along the identity path is (\ref{Fhidentity}). 
\newline\newline
In section \ref{MM} we review the non-unitary minimal models $\mathcal{M}_{2m-1,2}$, and discuss the scaling of their OPE coefficients in the large $m$ (or equivalently large negative central charge) limit for different choices of primary operators. The dominant OPE coefficients, which grow exponentially in $m$, stem from three primary operators whose conformal dimension is of order $m$. The resulting expression is given in (\ref{opegen}). The OPE coefficients are the building blocks of the sewing formula, which we use in section \ref{sec:sewing} to obtain large $m$ expressions (\ref{CFT_h}) for the genus $h\ge 2$ partition functions of $\mathcal{M}_{2m-1,2}$. The two-sphere partition function of minimal models is more subtle. We investigate it using the Coulomb gas formalism in section \ref{sec:SpherePart}. 
\newline\newline
In (\ref{Zgrav_intro}) only the identity operator of $\mathcal{M}_{2m-1,2}$ is coupled to gravity. As such, up to an integral over the moduli of the Riemann surface $\Sigma_h$, we can treat the matter, gravity and ghost contributions in (\ref{Zgrav_intro}) independently. The ghost partition function, discussed in section \ref{sec:ghost}, arises upon restricting to Weyl gauge. In this gauge the gravitational path integral (\ref{Zgrav_intro}) reduces to that of Liouville theory. We review the large $m$ limit of the sphere and torus partition function of Liouville theory in section \ref{sec:ST_L}. In section \ref{liouville_higherGenus} we study the large $m$ limit of Liouville theory on Riemann surfaces of genus $h\geq 2$, resulting in the expression (\ref{ZLgen}).  Section \ref{sec:comparison} is devoted to a comparison of normalisation independent ratios (\ref{ratio1}) stemming from the planar genus expansion of $\lim_{\text{d.sc.}} \log\mathcal{M}_N^{(m)}(\boldsymbol{\alpha}) $ and the genus expansion of $\mathcal{Z}^{(m)}_{\text{grav}}[\Lambda,\vartheta]$. 
\newline\newline
In section \ref{sec:dS} we end with a discussion on the relation of our results to the theory of a two-dimensional de Sitter universe. We provide hints for an entropic interpretation of $\mathcal{Z}^{(m)}_{\text{grav}}[\Lambda,\vartheta]$ and speculate on a Lorentzian interpretation of the fixed area constraint as an averaging over a complexified cosmological constant. 


\section{Multicritical matrix integrals at large $m$}\label{matrix}

In this section, we discuss some results for multicritical matrix integrals \cite{Kazakov:1989bc,Staudacher:1989fy,Moore:1991ir} at large $m$. As reviewed and explored in \cite{penguins}, these are given by integrals over an $N\times N$ Hermitian matrix $M$
\begin{equation}\label{MI}
\mathcal{M}^{(m)}_N(\boldsymbol{\alpha}) = \int_{\mathbb{R}^{N^2}} [\dd M] e^{- N \text{tr} V_m(M,\boldsymbol{\alpha})}~, \quad m = 2,3,\ldots~,
\end{equation}
where $V_m(M,\boldsymbol{\alpha})$ is the matrix polynomial
\begin{equation}\label{potentials}
V_m(M,\boldsymbol{\alpha}) = \sum_{n=1}^m \frac{1}{2 n} \alpha_{n} M^{2n}~,\quad \alpha_1=1~,
\end{equation}
and $\boldsymbol{\alpha} = (\alpha_2,\alpha_3,\ldots,\alpha_m)$ is a set of $(m-1)$ real valued coupling constants that are tuned to reside near the multicritical point denoted by $\boldsymbol{\alpha}^{(m)}_c$. Explicitly,
\begin{equation}\label{alpha_c}
{\alpha}^{(m)}_{n,c} \equiv (-1)^{n+1} \binom{m}{n} \frac{2n}{(4m)^n} B(n,1/2)~, \quad n=2,3,\ldots, m~,
\end{equation}
where $B(x,y)$ denotes the beta function. The 't Hooft genus expansion takes the following form
\begin{equation}\label{planar}
\mathcal{F}^{(m)}( \boldsymbol{\alpha} ) \equiv - \lim_{N\to \infty} \log \frac{\mathcal{M}^{(m)}_N(\boldsymbol{\alpha})}{\mathcal{M}^{(m)}_N(\bold{0})} = \sum_{h=0}^\infty N^{2-2h} \mathcal{F}_{h}^{(m)}  ( \boldsymbol{\alpha} )~,
\end{equation}
with $h$ denoting the genus. 

\subsection{The string equation for multiple paths}

Approaching $\boldsymbol{\alpha}^{(m)}_c$ from some point in coupling space must be defined with care \cite{penguins}, as the non-analytic behaviour of $\mathcal{F}_h^{(m)}( \boldsymbol{\alpha} )$ depends on the path. One possible set of paths leading to ($m-1$) distinct potentials in (\ref{potentials})  and consequently $(m-1)$ distinct non-analytic behaviours of $\mathcal{F}_h^{(m)}( \boldsymbol{\alpha} )$, can be parameterised as follows
\begin{equation}\label{paths}
\gamma_s^{(m)}(t) = \left(1, \alpha^{(m)}_{2,c} \, t^{s_2}, \alpha^{(m)}_{3,c} \, t^{s_3} ,\ldots , \alpha^{(m)}_{m,c} \, t^{s_{m}} \right)~, \quad\quad t \in [0,1]~.
\end{equation}
The general procedure to determine the $s_n$ is explained in \cite{penguins}, and some details are provided in appendix \ref{app:paths}. The multicritical point lies at $t=1$, and the non-analytic behaviour emerges as we approach $t = 1^-$. Of the paths $\gamma_s^{(m)}$, we will focus on the one with $s_n = n(n-1)/(m(m-1))$ which we refer to as the identity path
\begin{equation}\label{gammaid}
\gamma_{\text{id}}^{(m)}(t)= \begin{pmatrix}
\alpha_{2,c}^{(m)}t^{\frac{2}{m}\frac{1}{m-1}}\\
\vdots \\
\alpha_{n,c}^{(m)}t^{\frac{n}{m}\frac{n-1}{m-1}}\\
\vdots\\
\alpha_{m,c}^{(m)}t
\end{pmatrix}~.
\end{equation}
As shown in appendix \ref{app:paths}, in the planar limit, and taking $t = 1-\epsilon$ with $\epsilon \ll 1$, the identity path leads to the non-analytic behaviour
\begin{equation}\label{F0_path}
\mathcal{F}_{0}^{(m)}(\epsilon) = \frac{4}{(2m-3)(2m-1)(2m+1)}\epsilon^{m+\frac{1}{2}}~.
\end{equation}
{In section \ref{sec:gravity} we show that from a continuum perspective the identity path $\gamma_{\text{id}}^{(m)}(t)$ corresponds to coupling the identity operator of the minimal model $\mathcal{M}_{2m-1,2}$ to gravity, i.e. that $\epsilon$ is proportional to the cosmological constant $\Lambda$.}

Our goal at this stage is to evaluate the large $m$ behaviour of $\mathcal{M}^{(m)}_N(\boldsymbol{\alpha})$ along $\gamma_{\text{id}}^{(m)}(t)$. In particular, we will assess the large $m$ behaviour of the $\mathcal{F}_{h}^{(m)}(\boldsymbol{\alpha})$ in the expansion (\ref{planar}) for general genus $h\geq 0$. There are various methods to do so. Our approach follows that of \cite{Belavin:2008kv,Belavin:2010pj,Belavin:2010bs} which computes the $\mathcal{F}_{h}^{(m)}(\boldsymbol{\alpha})$ using what is known as the string equation \cite{Douglas:1989ve,Gross:1989vs,Brezin:1990rb}. In this approach, which can be derived starting from the orthogonal polynomial solution to the matrix integral as we review in appendix \ref{app:SE}, one expresses the $\mathcal{F}_{h}^{(m)}$ in terms of a general order $m$ polynomial
\begin{equation}\label{Polynomial}
\mathcal{P}_m(u) = u^m - \epsilon \, u^{m-2}  -   t_{m-3} u^{m-3} - \ldots - z~, 
\end{equation}
where the $t_n $ correspond to turning on the specific combination of couplings (parametrically close to $\boldsymbol{\alpha}_c^{(m)}$) leading to one of the $(m-1)$ types of non-analytic behaviours of $\mathcal{F}_{h}^{(m)}$ \cite{penguins}. For the identity path, we set $z = t_1 = \ldots =t_{m-3} = 0$. In particular, following and extending \cite{Belavin:2008kv,Belavin:2010pj,Belavin:2010bs} to $\mathcal{F}_{h=4}^{(m)}(\epsilon)$, we find 
\begin{eqnarray}\label{F_matrix}
\mathcal{F}_{0}^{(m)}(\epsilon) &=& \frac{1}{2}\int^{u^*} du   \mathcal{P}_m(u)^2 \approx \frac{1}{2} m^{-3} \epsilon^{m+\frac{1}{2}}+ \mathcal{O}(m^{-2})~, \\ \label{F1eps}
\mathcal{F}_{1}^{(m)}(\epsilon) &=&  \frac{1}{12} \log  \partial_u \mathcal{P}_m(u^*) = \frac{(m-1)}{24} \log \epsilon~, \\
\mathcal{F}_{2}^{(m)}(\epsilon) &=& \frac{1}{1440} \partial_u \left( 7  \frac{\partial^2_u \mathcal{P}_m(u^*)^2}{\partial_u \mathcal{P}_m(u^*)^4} - 5 \frac{\partial^3_u \mathcal{P}_m(u^*)}{\partial_u \mathcal{P}_m(u^*)^3}\right) \approx  -\frac{7}{576} m^{3} \epsilon^{-\left(m+\frac{1}{2}\right)}+ \mathcal{O}(m^2)~.
\end{eqnarray}
For $\mathcal{F}_{3}^{(m)}(\epsilon)$, we have a slightly lengthier (but no less interesting!) expression
\begin{align}\label{F3_matrix}
&\mathcal{F}_{3}^{(m)} (\epsilon)= -\frac{245}{2592}\frac{\partial_u^2 \mathcal{P}_m(u^*)^6}{\partial_u\mathcal{P}_m(u^*)^{10}} + \frac{193}{864}\frac{\partial_u^2 \mathcal{P}_m(u^*)^4\partial_u^3\mathcal{P}_m(u^*)}{\partial_u \mathcal{P}_m(u^*)^9} \cr
&- \frac{1}{1728}\frac{\partial_u^2\mathcal{P}_m(u^*)^2}{\partial_u\mathcal{P}_m(u^*)^8}\left(205\partial_{u}^3\mathcal{P}_m(u^*)^2+106 \partial_u^2 \mathcal{P}_m(u^*)\partial_{u}^4\mathcal{P}_m(u^*)\right)\cr
&+\frac{1}{288}\frac{1}{\partial_u\mathcal{P}_m(u^*)^7}\left(\frac{583}{252}{\partial_{u}^3\mathcal{P}_m(u^*)^3}+ \frac{1121}{105}{\partial_{u}^2\mathcal{P}_m(u^*)\partial_u^3\mathcal{P}_m(u^*)\partial_u^4\mathcal{P}_{m}(u^*)} + \frac{17}{5}{\partial_{u}^2\mathcal{P}_m(u^*)^2\partial_{u}^5\mathcal{P}_m(u^*)}\right)\cr
&- \frac{1}{\partial_u\mathcal{P}_m(u^*)^6}\left(\frac{607}{362880}\partial_{u}^4\mathcal{P}_m(u^*)^2+ \frac{503}{181440}\partial_{u}^3\mathcal{P}_m(u^*)\partial_{u}^5\mathcal{P}_m(u^*)+ \frac{77}{51840}\partial_{u}^2\mathcal{P}_m(u^*)\partial_{u}^6\mathcal{P}_m(u^*)\right) \cr
&+\frac{\partial_u^7\PP(u^*)}{10368\partial_u\PP(u^*)^5} \approx -\frac{1531}{80640} m^{6} \epsilon^{-2\left(m+\frac{1}{2}\right)}+ \mathcal{O}(m^5)~, 
\end{align}
with $\mathcal{P}_m(u^*) = 0$. We choose the positive root $u^* = \sqrt{\epsilon}$. Details and explicit expressions for 
\begin{equation}\label{F4eps}
\mathcal{F}_{4}^{(m)}(\epsilon) \approx - \frac{19016069}{174182400} m^{9}\epsilon^{-3\left(m+\frac{1}{2}\right)} + \mathcal{O}(m^8)~ 
\end{equation}
can be found in appendix \ref{app:SE}. From the above expressions it is natural to conjecture the general large $m$ dependence to be
\begin{equation}\label{Fhidentity}
\mathcal{F}_{h}^{(m)}(\epsilon) \approx f_h \, m^{3(h-1)}\epsilon^{-(h-1)\left(m+ \frac{1}{2}\right)} + \ldots~,\quad h\neq 1
\end{equation}
with $f_h$ independent of $m$. For genus $h=1$ we obtain (\ref{F1eps}) which exhibits a logarithmic non-analyticity in $\epsilon$ with a coefficient growing linearly in $m$.\footnote{A different approach to calculate the $\mathcal{F}_{h}^{(m)}(\epsilon) $ follows the tools of topological recursion, see e.g. \cite{Gregori:2021tvs}.} 

Finally, we note that setting $\epsilon= t_{m-3}= \ldots =t_1=0$ while keeping $z$ non-vanishing, the expressions (\ref{F_matrix})-(\ref{F4eps}) yield the free energy along the `minimal' path corresponding to $s_n= (n-1)$ in (\ref{paths}), and we recover the expressions presented in the appendix of \cite{Gross:1989vs}. {From a continuum perspective the minimal path corresponds to turning on a coupling for the operator of lowest conformal dimension in $\mathcal{M}_{2m-1,2}$.} For the minimal path we observe the leading large $m$ non-analyticity
\begin{equation}\label{Fhminimal}
\mathcal{F}_{0}^{(m)}(z) \approx \tilde{f}_0\, z^{2+\frac{1}{m}}~, \quad \mathcal{F}_{h}^{(m)}(z) \approx \tilde{f}_h \, m^{h-1}z^{-(h-1)\left(2+ \frac{1}{m}\right)} + \ldots~,\quad h\geq 2~,
\end{equation}
with $\tilde{f}_0$ and $\tilde{f}_h$ independent of $m$. For genus $h=1$ we obtain a logarithmic non-analyticity in $z$, whose coefficient is of order one \cite{Belavin:2010pj}. The other paths interpolate between the large $m$ scalings of the identity (\ref{Fhidentity}) and the minimal (\ref{Fhminimal}) path respectively. 

\subsection{Other quantities at large $m$}

To conclude this section we mention the large $m$ limit of two other important quantities of the multicritical matrix integral $\mathcal{M}^{(m)}_N(\boldsymbol{\alpha})$. In the large $m$ limit and upon tuning the couplings to the multicritical point (\ref{alpha_c}) the polynomial (\ref{potentials}) reduces to \cite{Ambjorn:2016lkl}
\begin{equation}
\lim_{m\rightarrow \infty}V_m\left(\lambda,\boldsymbol{\alpha}_c^{(m)}\right)= \frac{1}{2}\lambda^2\,_2F_2\left(1,1; \frac{3}{2},2; -\frac{\lambda^2}{4}\right)~.
\end{equation}
The large $m$ limit of the extremum of the eigenvalue density $ \rho^{(m)}_{\mathrm{ext}}(\lambda,\boldsymbol{\alpha})$ at the multicritical point (see appendix \ref{app:paths}) is given by
\begin{equation}
\lim_{m \to \infty} \rho^{(m)}_{\mathrm{ext}}\left(\lambda,\boldsymbol{\alpha}_c^{(m)}\right)=  \frac{1}{2\sqrt{\pi}}e^{-\lambda^2/4}~, \quad\quad \lambda \in \mathbb{R}~.
\end{equation}
In contrast to the Wigner semicircle distribution, $\rho^{(\infty)}_{\mathrm{ext}}(\lambda,\boldsymbol{\alpha}_c^{(m)})$ takes values all the way to infinity and being a pure Gaussian is infinitely differentiable (see also \cite{Tierz:2001kv}). 

It is also of interest to consider the large $m$ limit of the expectation value of the macroscopic loop operator $W^{(m)}_\ell \equiv \text{Tr} e^{-\ell M}$, which is associated to the manifolds with boundaries in the planar expansion. To leading order in the planar expansion, one finds the large $m$ expansion
\begin{equation}
\lim_{m\to\infty}\left\langle W_\ell^{(m)}\left(\boldsymbol{\alpha}_c^{(m)}\right)\right\rangle = e^{\ell^2}~.
\end{equation}
The resolvent can be obtained from $W^{(m)}_\ell$ by a Laplace transform. 
\begin{center} *** \end{center}
We thus see that the family of matrix integrals $\mathcal{M}^{(m)}_N(\boldsymbol{\alpha})$ in the planar limit permits a rich large $m$ expansion. From a diagrammatic perspective \cite{penguins}, this data characterises the behaviour of multi-vertex planar diagrams decorated with vertices emanating an arbitrary numbers of edges. Given this large $m$ expansion, it is interesting to note that we can extract various `pure' functions of $m$ from $\mathcal{F}^{(m)}( \boldsymbol{\alpha} )$ in (\ref{planar}) that are independent of $\epsilon$ and $N$. For instance, near the end point of the identity path
\begin{align}\label{ratio1}\nonumber
\mathcal{F}_0^{(m)}(\epsilon)\times \mathcal{F}_2^{(m)}(\epsilon) &= \frac{1}{1440}\frac{(12-142 m+201m^2- 70m^3)}{(2m-3)(2m-1)(2m+1)}\approx  -\frac{7}{1152}+ \frac{1}{120}\frac{1}{m}- \frac{31}{23040}\frac{1}{m^2}+\ldots~,\\ \nonumber
\left(\mathcal{F}_0^{(m)}(\epsilon)\right)^2\times \mathcal{F}_3^{(m)}(\epsilon) &\approx - \frac{1531}{322560}+ \frac{7933}{967680}\frac{1}{m}- \frac{9493}{2903040}\frac{1}{m^2}+ \ldots~,\\ \nonumber
\frac{\mathcal{F}_3^{(m)}(\epsilon)}{\left(\mathcal{F}_2^{(m)}(\epsilon)\right)^2} &\approx - \frac{220464}{1715} -\frac{7837104}{60025}\frac{1}{m} - \frac{310874008}{2100875}\frac{1}{m^2}+ \ldots~,\\
\frac{\mathcal{F}_4^{(m)}(\epsilon)}{ \mathcal{F}_2^{(m)}(\epsilon) \times \mathcal{F}_3^{(m)}(\epsilon) } & \approx -\frac{76064276}{160755}-\frac{12247926130684}{25842170025}\frac{1}{m}-\frac{243356317673253649}{461584226929875}\frac{1}{m^2}+ \ldots~.
\end{align}
One can also compute quantities that are insensitive to the overall normalisation of $\mathcal{F}^{(m)}$. For instance
\begin{equation}\label{ratio2}
\frac{\mathcal{F}_0^{(m)}(\epsilon) \times \left( \mathcal{F}_3^{(m)}(\epsilon) \right)^2}{\left( \mathcal{F}_2^{(m)}(\epsilon) \right)^3} \approx -\frac{42191298}{420175}-\frac{974469252}{14706125}\frac{1}{m}- \frac{79528573029}{1029428750}\frac{1}{m^2} + \ldots~.
\end{equation}
Recalling the conjectural relation \cite{Kazakov:1989bc,Staudacher:1989fy} between multicritical matrix integrals and two-dimensional quantum gravity coupled to non-unitary minimal model $\mathcal{M}_{2m-1,2}$, it is natural to ask about analogous large $m$ limits, and pure functions of $m$, in the continuum picture. 


\section{$\mathcal{M}_{2m-1,2}$ minimal models at large $m$}\label{MM}

In this section, we discuss some results for the $\mathcal{M}_{2m-1,2}$ family of minimal models at large $m$. It is worth recalling \cite{Belavin:1984vu} that these minimal models have $(m-1)$ Virasoro primaries, $\mathcal{O}_{1,s}$ with $s=1,2,\ldots,m-1$, whose holomorphic weights are given by
\begin{equation}\label{dimensions}
\Delta_{1,s} = \frac{(2m-1-2s)^2 - (2m-3)^2}{8(2m-1)}~,
\end{equation} 
and whose central charge is 
\begin{equation}\label{cm}
c_m = 1-\frac{3(3-2m)^2}{2m-1} \approx -6 m + 16 + \ldots~.
\end{equation}
We note that conformal dimensions (\ref{dimensions}) are non-positive, and $c_m< 0$, reflecting the non-unitary nature of $\mathcal{M}_{2m-1,2}$ on a rigid background.
The identity operator $\mathcal{O}_{1,1}$ has the highest conformal dimension. The operator of minimal weight is $\mathcal{O}_{1,m-1}$ with 
\begin{equation}\label{Delta_min}
\Delta_{1,m-1} = \frac{(m-1)(m-2)}{2(1-2m)} \approx -\frac{m}{4} + \frac{5}{8} + \ldots~.
\end{equation}
For large $m$, we note that $\Delta_{1,m-1}  \approx 24 c_m$. The remaining ingredient defining the models are the OPE coefficients. These were first computed by Dotsenko and Fateev using the Coulomb Gas method \cite{Dotsenko:1985hi,Dotsenko:1984ad,Dotsenko:1986ca}. For the $\mathcal{M}_{2m-1,2}$ minimal models they read
\begin{equation}\label{OPE_DF}
C_{s_1 ,s_2}^{s_3} \equiv \left(\frac{a_{s_1}a_{s_2}}{a_{s_3}}\right)^{\frac{1}{2}}\times D_{s_1 ,s_2}^{s_3}~,
\end{equation}
where
\begin{equation}\label{def:Ca}
D_{s_1, s_2}^{s_3} \equiv \prod_{i=1}^{k-1}\frac{\gamma(\rho i)}{\gamma\left(\rho (s_1-i)\right)\gamma\left(\rho (s_2-i)\right)\gamma\left(2- \rho (s_3+i)\right)}~,\quad
a_s \equiv \prod_{i=1}^{s-1}\frac{\gamma\left(\rho i\right)}{\gamma\left(-1+\rho (1+i)\right)}~,
\end{equation}
and
\begin{equation}
k\equiv \frac{1}{2}(s_1+s_2-s_3+1)~,\quad \rho \equiv \frac{2}{2m-1}~,\quad  \gamma(z)\equiv \frac{\Gamma(z)}{\Gamma(1-z)}~.
\end{equation}
In order to select the relevant indices, we must recall the fusion rules of $\mathcal{M}_{2m-1,2}$ which read
\begin{equation}\label{fusiongen}
\mathcal{O}_{1,s_1} \times \mathcal{O}_{1,s_2} = \sum_{s_3= 1+|s_1-s_2| \,\, \text{mod}~ 2}^{\text{min}(s_1+s_2-1,4m-3-s_1-s_2)} \mathcal{O}_{1,s_3}~.
\end{equation}
Whereas (\ref{OPE_DF}) yields a non-vanishing result for general values of $s_1,s_2$ and $s_3$, one should only consider those values of $s_1,s_2$ and $s_3$ in (\ref{OPE_DF}) following from the fusion rules (\ref{fusiongen}). We note that {only} the cases where either all of the $s_i$ are odd or  if up to permutations they are of the form even-even-odd does the OPE coefficient reflect a physical three-point function in $\mathcal{M}_{2m-1,2}$. It is also worth recalling the operator relation $\mathcal{O}_{1,s} = \mathcal{O}_{1,2m-1-s}$. 

We normalise all operators such that their two-point function has unit coefficient. In this normalisation, some of the $C_{s_1,s_2}^{s_3}$ are imaginary, indicating the non-unitarity of the $\mathcal{M}_{2m-1,2}$ models. As an example, we can compute the OPE coefficient $C^3_{2,2}$ of the only non-trivial operator for the $\mathcal{M}_{5,2}$ Lee-Yang model \cite{Cardy:1985yy} which reads $C^3_{2,2} =  1.911... \times i$ and is pure imaginary. For the minimal model $\mathcal{M}_{7,2}$ we find both purely imaginary as well as purely real OPE coefficients. As an example we find $C_{3,3}^3= 6.019...$ and $C_{3,3}^5= 4.592... \times i$. As a final remark we note that from the expression (\ref{OPE_DF}) it is immediate that the reality properties are tied to the sign of $a_s$, for which we find
\begin{equation}
a_s \sim (-1)^{s+1}~,\quad s \leq m-1~,\quad a_s \sim (-1)^s~,\quad s\geq m~.
\end{equation}

\subsection{OPE coefficients at large $m$}
In this section we discuss the large $m$ behaviour of the OPE coefficients (\ref{OPE_DF}) for the non-unitary minimal models $\mathcal{M}_{2m-1,2}$. An analysis of the asymptotic behaviour of the OPE coefficients for unitary compact two-dimensional conformal field theories has been performed in \cite{Maloney,Fitzpatrick:2014vua}.

To keep the presentation simple, we will discuss here some simple cases while discussing the more general and richer situation in appendix \ref{minimal}. 
\vspace{3mm} \\
{\textbf{Three minimal operators $\mathcal{O}_{1,m-1}$.}} We begin by recalling the fusion rule
\begin{equation}
\mathcal{O}_{1,m-1} \times \mathcal{O}_{1,m-1} = \sum_{l=0}^{m-2}\mathcal{O}_{1,2l+1} = \mathcal{O}_{1,1} + \mathcal{O}_{1,3} + \ldots + \mathcal{O}_{1,2m-3}~,
\end{equation}
and the relation $\mathcal{O}_{1,s} = \mathcal{O}_{1,2m-1-s}$, which implies $\mathcal{O}_{1,m-1} = \mathcal{O}_{1,m}$.  These are the operators of lowest conformal dimension in $\mathcal{M}_{2m-1,2}$ with conformal dimension (\ref{Delta_min}).
For even values of $m$, the fusion rule of $\mathcal{O}_{1,m-1}$ with itself contains $\mathcal{O}_{1,m-1}$. For odd values of $m$, it contains $\mathcal{O}_{1,m}$. Thus, for even values of $m$ we are interested in the OPE coefficient $C_{m-1,m-1}^{m-1}$ while for odd values of $m$ we are interested in the OPE coefficient $C_{m-1,m-1}^{m}$. For the sake of simplicity we focus on the case where $m$ is even, i.e. $C_{m-1,m-1}^{m-1}$.  
Explicitly \cite{Dotsenko:1986ca,Belavin:2003pu},
\begin{equation}\label{cmmm}
C_{m-1,m-1}^{m-1} = 
\prod _{k=1}^{m-2} \frac{\gamma \left(\frac{2 k}{2 m-1}\right)^{\frac{1}{2}}}{ \gamma \left(-1+\frac{2 (k+1)}{2 m-1}\right)^{\frac{1}{2}}} \, \times \prod _{k=1}^{\frac{m-2}{2}} \frac{\gamma \left(\frac{2 k}{2 m-1}\right)}{\gamma\left(\rho (m-1-k)\right)^2\gamma\left(2- \rho (m-1+k)\right)}~.
\end{equation}
We would like to estimate the above expression in the large $m$ limit. To do so, we consider the logarithm of $C_{m-1,m-1}^{m-1}$ and approximate the resulting sum using the Euler-Maclaurin approximation \cite{wiki}. 
Keeping the two leading terms of the Euler-Maclaurin approximation, we arrive at the large $m$ expression
\begin{equation}\label{opelargem}
\lim_{m\to \infty} C_{m-1,m-1}^{m-1} \approx v_{1,1}^1 \, m^{-3/2} \, e^{\nu m}~, \quad\quad \nu \equiv 12  \log A-1-\frac{1}{3}  \log 2>0~,
\end{equation}
where $A=1.28\ldots$ is the Glaisher constant and $v_{1,1}^1$ is independent of $m$. To verify the above, we must assess the size of the errors in the Euler-Maclaurin formula. Fortunately, these are well understood and yield terms which are all subleading at large $m$. More generally, as shown in appendix \ref{minimal}, given three near minimal weight operators of the type $\mathcal{O}_{1,m-r}$ with $r$ independent of $m$ we have
\begin{equation}\label{opegen}
\lim_{m\to \infty} C_{m-r_1,m-r_2}^{m-r_3} \approx v_{r_1,r_2}^{r_3} \, m^{-3/2} \, e^{\nu m}~. 
\end{equation}
Interestingly, the $r$ dependence only appears at the level of the $m$ independent pre-factor, which is either pure real or pure imaginary. Recalling (\ref{Delta_min}) and (\ref{cm}) we observe that the conformal dimensions of  the $\mathcal{O}_{1,m-r}$ with $r \sim\mathcal{O}(1)$ scale like $c_m$ at large $m$. In other words, (\ref{opegen}) describes the behaviour of the OPE coefficients for three primary operators of $\mathcal{M}_{2m-1,2}$ in the limit of large and negative $c_m$ with $\Delta_{1,m-r_i}/c_m \sim \mathcal{O}(1)$.

Some insight into the specific form of (\ref{opelargem}) comes from considering the large $m$ limit of the OPE coefficient of a composite operator $
\phi^m$ in a generalised free field theory (see for example \cite{Belin:2017nze}). The structure of the OPE coefficient at large $m$ in the generalised free theory takes the form $\sim m^\beta e^{\nu m}$ but with different values of $\beta$ and $\nu$ than those in  (\ref{opelargem}). In this case, the exponential growth stems from a combinatorial factor.\footnote{Perhaps relatedly, in the Landau-Ginzburg description \cite{amoruso} of the $\mathcal{M}_{2m-1,2}$ minimal models the minimal operators are given by composite operators $\varphi^{m-2}$ built from the Landau-Ginzburg field $\varphi$.} 
\vspace{3mm} \\
\textbf{Three light operators $\mathcal{O}_{1,3}$.} We now consider the large $m$ limit for the OPE coefficient $C^3_{33}$ of the three light operators $\mathcal{O}_{1,3}$. In this case we have
\begin{equation}
C^3_{3,3} =2^{-\frac{5}{2}-\frac{8}{2m-1}} \times  (2m-5)\frac{\gamma\left(\frac{1}{2}- \frac{2}{2m-1}\right)}{\gamma\left(\frac{3}{2}- \frac{4}{2m-1}\right)} \frac{\gamma\left(2- \frac{6}{2m-1}\right)^{\frac{1}{2}}}{(2m+1)^{\frac{1}{2}}}\frac{\Gamma\left(\frac{2}{2m-1}\right)^{\frac{1}{2}}}{\left(-\Gamma\left(-1-\frac{2}{2m-1}\right)\right)^{\frac{1}{2}}}~.
\end{equation}
At large $m$ we have that
\begin{equation}
C^3_{3,3} = -\sqrt{3}-\frac{3 \sqrt{3}}{m}-\frac{31 \sqrt{3}}{2 m^2}+\frac{3 \sqrt{3} (32 \zeta (3)-105)}{4 m^3}+\frac{27 \sqrt{3} (32 \zeta (3)-117)}{8 m^4}+ \ldots~.
\end{equation}
We note that the above expression is independent of $m$ at large $m$. Similarly, we find that $C_{r_1,r_2}^{r_3} \sim\mathcal{O}(1)$ if $r_1,r_2$ and $r_3$ do not scale with $m$ in the large $m$ limit.
\vspace{3mm} \\
\textbf{Two minimal and one light operator.} Finally we consider the large $m$ limit of two near minimal operators and the light operator $\mathcal{O}_{1,3}$. At large $m$ we find that 
\begin{equation}
C^{m-1}_{m-1,3} \approx \text{const}\times (-1)^m m^2 + \ldots 
\end{equation}
We note the absence of any exponential growth in $m$. The absence of any exponential growth holds true for the more general case of minimal-minimal-light OPE coefficients.
\vspace{3mm} \\
\textbf{Two light and one minimal operator.}  We can also ask about the large $m$ behaviour of OPE coefficients associated to two operators of order one conformal dimension, and one near minimal operator. Inspection of the general fusion rules (\ref{fusiongen}) of the $\mathcal{M}_{2m-1,2}$ minimal models shows that such OPE coefficients vanish. 

\subsection{Higher genus partition function at large $m$}\label{sec:sewing}

At this stage, we can use the large $m$ expressions of the OPE coefficients to estimate the partition function $Z_{\text{CFT}}^{(m)}[\Sigma_h]$ of $\mathcal{M}_{2m-1,2}$ on a compact Riemann surface $\Sigma_h$ of genus $h$. We take the Ricci scalar to be $R=-2(h-1)/\upsilon$, such that the area is $4\pi \upsilon$. To do so we employ the sewing formula \cite{Friedan:1986ua,Vafa:1988pw,Sonoda:1988mf,Sonoda:1988fq}. We follow the treatment in \cite{Sen:1990bt}. 

To write down the sewing formula for $h\ge 2$, it is convenient to first arrange $\Sigma_h$ in the form of $2h-2$ two-spheres each with three-punctures connected via $3h-3$ cylindrical tubes. For a general two-dimensional conformal field theory the sewed partition function reads 
\begin{equation}\label{sewing}
Z_{\text{CFT}}[\Sigma_h] = \left( \frac{\upsilon}{\upsilon_0} \right)^{-\frac{(h-1)(c+\bar{c})}{12}} \left( Z_{\text{CFT}}[S^2] \right)^{2h-2} \sum_{i_1,\ldots,i_{3h-3}} \prod_{l=1}^{3h-3} \tau_l^{\Delta_{i_l}-\frac{c}{24}}  \bar{\tau}_l^{\bar{\Delta}_{i_l}-\frac{\bar{c}}{24}} \prod_{n=1}^{2h-2}C^{{k_n}}_{{i_n}  {j_n}}~.
\end{equation}
In the above expression, the sum runs over all local operators (with respective holomorphic and anti-holomorphic conformal weights $\Delta_i$ and $\bar{\Delta}_i$) in the conformal field theory of holomorphic and anti-holomorphic central charge $c$ and $\bar{c}$. The index $l$ labels the $3h-3$ tubes connecting the punctured two-spheres. The moduli $(\tau_l,\bar{\tau}_l)$ are complex numbers residing within the unit disk.  The $C^{{k_n}}_{{i_n}  {j_n}}$
are the OPE coefficients associated to the three-operators on the $n^{\text{th}}$ three-punctured two-sphere, {\color{black}and the collection of $2h-2$ triplets $\{  i_1, j_1, k_1 ; \ldots ;   i_{2h-2}, j_{2h-2} , k_{2h-2} \}$ denote the operators on each punctured sphere. Since each tube connects one of these operators to another on a distinct two-sphere obtained by evolving along the length of the tube, we have in total $3h-3$ operators to sum over, denoted by $i_1$, \ldots $i_{3h-3}$ in the overall sum in (\ref{sewing}).} The $\upsilon$ dependent pre-factor in (\ref{sewing}) stems from the conformal anomaly, with $\upsilon_0$ being a reference area. Finally, we have reinstated the dependence of the two-sphere partition function $Z_{\text{CFT}}[S^2]$ which has been normalised\footnote{It is worth emphasising that the area of the $S^2$ is taken to be some fixed reference area. As such there is no contribtion from $Z_{\text{CFT}}[S^2]$ to the $\upsilon$ dependence in (\ref{sewing}) which results from the conformal anomaly.} to one in \cite{Sen:1990bt}.  {As a concrete example with all the indices written out, the genus-two partition function is given
\begin{equation}
Z_{\text{CFT}}^{(m)}[\Sigma_2] =  \left( \frac{\upsilon}{\upsilon_0} \right)^{-\frac{c_m}{6}} \left( Z_{\text{CFT}}^{(m)}[S^2] \right)^{2} \sum_{i_1,i_2,i_{3}}  \left| C^{{i_3}}_{{i_1} {i_2}}\right|^2 \times \prod_{l=1}^{3} \left( \tau_l \bar{\tau}_l \right)^{{\Delta}_{i_l}-\frac{{c}_{m}}{24}}~.
\end{equation}
Here the genus two surface is built by gluing two three-puncture two-spheres (see figure \ref{fig:sewing}), and we have assumed a diagonal basis of operators.

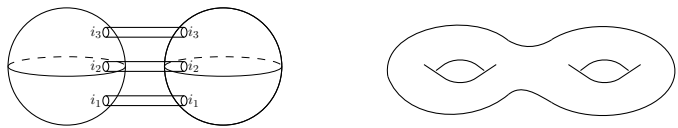
\begin{figure}[H]
\begin{center}
\begin{tikzpicture}[scale=1.3]
\node[scale=.5] at (-1.3,-.35)   {$i_1$};
\node[scale=.5] at (-1.3,0)   {$i_2$};
\node[scale=.5] at (-1.3,.35)   {$i_3$};
\node[scale=.5] at (-.3,-.35)   {$i_1$};
\node[scale=.5] at (-.3,0)   {$i_2$};
\node[scale=.5] at (-.3,.35)   {$i_3$};
\draw (-1.2,-.35) circle (.03cm and 0.05cm);
\draw (-.4,-.35) circle (.03cm and 0.05cm);
\draw (-1.2,.35) circle (.03cm and 0.05cm);
\draw (-.4,.35) circle (.03cm and 0.05cm);
\draw (-1.2,0) circle (.03cm and 0.05cm);
\draw (-.4,0) circle (.03cm and 0.05cm);
  \draw[line width=.4] (-1.6,0) circle (.6cm);
  \draw[line width=.3] (-2.2,0) arc (180:360:.6 and 0.1);
  \draw[dashed,line width=.3] (-2.2,0) arc (0:180:-.6 and 0.1);
  \draw (0,0) circle (.6cm);
  \draw (0,0) circle (.6cm);
  \draw[line width=.3] (-.6,0) arc (180:360:.6 and 0.1);
  \draw[dashed,line width=.3] (-.6,0) arc (0:180:-.6 and 0.1);
  \draw (-1.2,.4) -- (-.4,.4);
    \draw (-1.2,.3) -- (-.4,.3);
      \draw (-1.2,-.4) -- (-.4,-.4);
    \draw (-1.2,-.3) -- (-.4,-.3);
       \draw (-1.2,.05) -- (-.4,.05);
    \draw (-1.2,-.05) -- (-.4,-.05);   
\end{tikzpicture}
\quad\quad\quad
\begin{tikzpicture}[rotate= 180,scale=1.6]
\draw[] (-1.8607,-0.356) to[out=30,in=150] (-1.5,-0.35);
\draw[] (-1.8605,0.1656) to[out=-30,in=220] (-1.5,0.1);
\draw[] (-1.8605,0.1656) arc (45:315:.6 and 0.37);
\draw[] (-1.5,0.1) to[out=40,in=110] (-.5,-.05);
\draw[] (-1.5,-0.35) to[out=-40,in=-80] (-.5,-.05);
\draw[rounded corners=8pt] (-1.3,-.1)--(-1.1,-.25)--(-.9,-.1);
\draw[rounded corners=7pt] (-.8,-.15)--(-1.1,.05)--(-1.4,-.15);

\draw[rounded corners=8pt] (-2.5,-.1)--(-2.3,-.25)--(-2.1,-.1);
\draw[rounded corners=7pt] (-2,-.15)--(-2.3,.05)--(-2.6,-.15);
\end{tikzpicture}
\end{center}
\caption{Sewing two three-punctured spheres yields a genus two surface.}
\label{fig:sewing}
\end{figure}
We now apply the sewing formula (\ref{sewing}) to $\mathcal{M}_{2m-1,2}$ at large $m$. The basic property we will exploit is that the OPE coefficients  (\ref{opegen}) of three near minimal weight operators are exponentially large in $m$, while OPE coefficients involving at least one operator which is not near minimal weight are not. Consequently, the sum (\ref{sewing}) is dominated by the OPE coefficients $C_{m-r_1,m-r_2}^{m-r_3} $ (\ref{opegen}) and we can approximate it by
\begin{equation}\label{CFT_h}
Z_{\text{CFT}}^{(m)}[\Sigma_h] \approx \mathcal{N}(\tau_l,\bar{\tau}_l) \, \left( \frac{\upsilon}{\upsilon_0} \right)^{m(h-1)} \left( Z_{\text{CFT}}^{(m)}[S^2] \right)^{2(h-1)} m^{-3(h-1)} \, e^{2\nu m (h-1)}~,\quad h\geq 2~.
\end{equation}
The normalisation factor $\mathcal{N}(\tau_l,\bar{\tau}_l)$, which is independent of $m$ but dependent on the moduli, incorporates the effect of summing over $r_1$, $r_2$, and $r_3$ as well as the Virasoro descendants (the descendant states should have a normalised two-point function). We note that the dependence on the moduli in (\ref{sewing}) for near minimal operators is approximately $\sim (\tau_l \bar{\tau}_l)^{-1/24}$ and hence independent of $m$ at large $m$. Interestingly, at low genera the sewing formula is real valued irrespective of the reality properties of the $C^{s_3}_{s_1,s_2}$.

Although it would be interesting to understand the subleading structure of $Z_{\text{CFT}}^{(m)}[\Sigma_h]$, a thorough analysis of this is at this stage beyond the scope of our presentation.

\subsection{Remarks on the $\mathcal{M}_{2m-1,2}$ sphere partition function}\label{sec:SpherePart}

The sewing formula (\ref{sewing}) yields the CFT partition function of $\mathcal{M}_{2m-1,2}$ on a Riemann surface $\Sigma_h$ of genus $h\geq 2$. 
The partition function $Z^{(m)}_{\text{CFT}}[S^2]$ on a sphere with metric
\begin{equation}\label{S2metric}
\dd\tilde{s}^2 = \upsilon \left(\dd\theta^2 + \sin^2\theta \dd\phi^2 \right) =\frac{  4\upsilon  \dd z \dd\bar{z}}{(1+z\bar{z})^2}~,
\end{equation}
has general form given by
\begin{equation}
Z^{(m)}_{\text{CFT}}[S^2] = \left( \frac{\upsilon}{\upsilon_0} \right)^{\frac{c_m}{6}} \zeta_m~.
\end{equation}
Somewhat curiously \cite{Zamolodchikov:2001dz}, the computation of $Z^{(m)}_{\text{CFT}}[S^2]$ turns out to be more subtle. The pre-factor $\zeta_m$ is ambiguous due to the presence of the Euler character local counterterm. Nonetheless, once a regularisation procedure is fixed, one can unambiguously compare $\zeta_m$ for different values of $m$ or $Z^{(m)}_{\text{CFT}}[S^2] $ to other genera. Below we discuss an attempt toward computing this quantity using considerations from the Coulomb gas formalism. Unfortunately our results are inconclusive. 



We now review the main concepts of this formalism, for a detailed discussion we refer e.g. to \cite{Kapec:2020xaj}. Roughly speaking the Coulomb gas formalism is a machinery that produces data for minimal models by employing free field theory techniques in the presence of a background $U(1)$ charge. For the minimal models $\mathcal{M}_{2m-1,2}$, the main technical object is a path integral
\begin{equation}\label{CG_PI}
Z^{(m)}_{\text{CG}}[S^2;x_i] = \int [\mathcal{D}\varphi]e^{-\frac{1}{4\pi}\int_{S^2} \dd^2 x \sqrt{\tilde{g}}\left(\tilde{g}^{ij}\partial_i\varphi\partial_j \varphi+ QR[\tilde{g}]\varphi+ 4\pi \mu_- e^{2i\alpha_-\varphi}\right)} \prod_{i=1}^n \mathcal{V}_{\alpha_i}(x_i)~,
\end{equation}
with $Q\equiv  i(\alpha_++\alpha_-)$ and the $\mathcal{V}_\alpha = e^{2 i \alpha \varphi}$ of weight $\Delta_\alpha = \bar{\Delta}_\alpha = \alpha(Q-\alpha)$ are vertex operators whose $\alpha$ are judiciously chosen to match the operator content of the minimal model. The $\alpha_{\pm}$ are screening charges 
\begin{equation}
\alpha_+ \equiv \sqrt{\frac{2m-1}{2}}~,\quad \alpha_- \equiv -\sqrt{\frac{2}{2m-1}}~.
\end{equation}
The field $\varphi$ is a compact scalar with periodicity $\varphi \sim \varphi+2\pi\sqrt{2(2m-1)}$.

In the absence of any insertions, it is tempting to argue (along the lines of \cite{Distler:1988jt,David:1988hj}) using a shift $\varphi\rightarrow \varphi + i/\alpha_-\log  \upsilon \mu_-$ that $Z^{(m)}_{\text{CG}}[S^2]$  produces the correct sphere anomaly $\upsilon^{\frac{c_m}{6}}$.
However, upon splitting $\varphi$ into a constant piece $\varphi_0$ and a non-constant piece \cite{Goulian:1990qr}, we note that the path integral over $\varphi_0$ leads to a vanishing result for $Z^{(m)}_{\text{CG}}[S^2]$.
A minimal way to obtain a non-vanishing sphere path integral is to insert a single integrated operator 
\begin{equation}\label{Zreflected}
\tilde{Z}^{(m)}_{\text{CG}}[S^2] = \int [\mathcal{D}\varphi]e^{-\frac{1}{4\pi}\int_{S^2} \dd^2 x \sqrt{\tilde{g}}\left(\tilde{g}^{ij}\partial_i\varphi\partial_j \varphi+ QR[\tilde{g}]\varphi+ 4\pi \mu_- e^{2i\alpha_-\varphi} \right)} \, \int_{S^2} \dd^2 x \sqrt{\tilde{g}} \, \mathcal{V}_{\alpha_+ + \alpha_-}(x)~,
\end{equation}
that cancels the background charge. The parameter $\mu_-$ can be treated perturbatively due to the boundedness of $e^{2i\alpha_-\varphi}$. The vertex operator $\mathcal{V}_{\alpha_+ + \alpha_-}$ has vanishing conformal weight and is the `reflection' of the identity operator $\mathcal{V}_{0}$. 
Interestingly, at large $m$ (\ref{Zreflected}) permits a semiclassical expansion but,
although non-vanishing, the $\upsilon$ dependence of $\tilde{Z}^{(m)}_{\text{CG}}[S^2]$ no longer agrees with that of $\mathcal{M}_{2m-1,2}$.

Another potentially interesting avenue to compute ${Z}^{(m)}[S^2]$  follows the Landau-Ginzburg description of minimal models. The challenge here is to deal with strongly coupled theories whose scalar potential is either imaginary or unbounded \cite{amoruso} for the minimal models of interest. Perhaps the large $m$ limit leads to interesting simplifications.

\section{$\mathcal{M}_{2m-1,2}$ coupled to gravity at large $m$}\label{sec:gravity}

In this section, we discuss two-dimensional gravity coupled to the $\mathcal{M}_{2m-1,2}$ family of minimal models at large $m$. The gravitational path integral is given by
\begin{equation}\label{Zgrav0}
\mathcal{Z}^{(m)}_{\text{grav}}[\Lambda,\vartheta] \equiv \sum_{h=0}^\infty e^{\vartheta \chi_h}\mathcal{Z}^{(m)}_{h}[\Lambda]~, \quad \mathcal{Z}^{(m)}_{h}[\Lambda] = \int [\mathcal{D} g] e^{-\Lambda \int_{\Sigma_h} \dd^2 x\sqrt{g}} \, Z_{\text{CFT}}^{(m)}[g_{ij} ;\Sigma_h]~,
\end{equation}
where $\chi_h = 2-2h$ is the Euler character of the compact Riemann surface $\Sigma_h$ of genus $h$. We consider the problem in the Weyl gauge, whereby the physical metric is expressed as
\begin{equation}\label{physmetric}
g_{ij} = e^{2b \varphi} \tilde{g}_{ij}~.
\end{equation}
Here $\tilde{g}_{ij}$ is a fixed reference `fiducial' metric which for genus zero we take to be (\ref{S2metric}). For general details on the subject of two-dimensional quantum gravity, we refer to the reviews \cite{Ginsparg:1993is,DiFrancesco:1993cyw,musings,Klebanov:1991qa}. In the Weyl gauge, the gravitational path integral $\mathcal{Z}^{(m)}_{h}[\Lambda]$ maps to that of Liouville conformal field theory \cite{David:1988hj,Distler:1988jt}. The action for Liouville theory is
\begin{equation}\label{SL}
S^{(m)}_L[\varphi] = \frac{1}{4\pi} \int_{\Sigma_h} \dd^2 x \sqrt{\tilde{g}} \left( \tilde{g}^{ab} \partial_a \varphi \partial_b \varphi + Q {R}[\tilde{g}] \varphi + 4\pi \Lambda e^{2b\varphi} \right)~,
\end{equation}
with $Q = b+1/b$ and $b = \sqrt{2/(2 m-1)}$. The Liouville central charge is given by 
\begin{equation}\label{cL}
c_L = 1 + 6 Q^2 = 26-c_m = 6 m+10+\frac{12}{2 m-1}~,
\end{equation}
while exponential operators $\mathcal{O}_\alpha \equiv e^{2\alpha \varphi}$ have scaling dimension $\Delta_\alpha = \bar{\Delta}_\alpha = \alpha(Q-\alpha)$. In the large $m$ limit, we have that $c_L \approx 6 m$ is large and positive. From the Liouville theory perspective, the object of interest will be the path integral on $\Sigma_h$ 
\begin{equation}
Z^{(m)}_L[\Lambda;\Sigma_h] = \int \left[\mathcal{D} \varphi \right]  e^{-S^{(m)}_L[\varphi]}~,
\end{equation}
where care must be taken in dividing by the residual gauge group upon fixing the Weyl gauge for $h=0$ and $h=1$.\footnote{Our choice of measure over the space of fields stems from the flat metric in field space
\begin{equation}
\dd s^2 =  \int_{\Sigma_h} \dd^2 x \sqrt{\tilde{g}} \, \delta \varphi(x)^2~,
\end{equation}
which is local with respect to the fiducial metric $\tilde{g}_{ij}$. We normalise our path integral such that
\begin{equation}
1 = \int [\mathcal{D}\varphi] \exp \left( {-\Lambda_{\text{uv}} \int_{\Sigma_h} \dd^2x \sqrt{\tilde{g}} \varphi(x)^2} \right)~,
\end{equation}
where $\Lambda_{\text{uv}}$ has dimensions of 1/length$^2$.
For $h=0$ we must also divide the path integral by the volume of $SL(2,\mathbb{C})$, while for $h=1$ we must divide by the volume of $U(1)\times U(1)$. A treatment of these residual volumes is discussed in \cite{beatrix}.
}
Thus, we are interested in computing 
\begin{equation}\label{Zgrav}
\mathcal{Z}^{(m)}_{h}[\Lambda]= \int_{\mathcal{F}_h} \left[\dd\mu_h \right] \, Z^{(m)}_L[\Lambda;\Sigma_h] \, Z_{\text{ghost}}[\Sigma_h]   \, Z^{(m)}_{\text{CFT}}[\Sigma_h]~,
\end{equation}
where $Z^{(m)}_{\text{CFT}}[\Sigma_h]$ and $Z_{\text{ghost}}[\Sigma_h]$ are the matter and $\mathfrak{b}\mathfrak{c}$-ghost CFT partition functions evaluated on the fiducial metric $\tilde{g}_{ij}$, and $\dd \mu_h$ is a measure over the moduli space $\mathcal{F}_h$ of $\Sigma_h$. We can treat the gravity, ghost, and matter partition function in (\ref{Zgrav}) independently, up to the integral over moduli, because we have set the couplings between non-identity operators in $\mathcal{M}_{2m-1,2}$ and gravity to zero. The dependence on the Weyl factor $\varphi$ stemming from matter and ghost CFT conformal anomaly is subsumed into the Liouville action $S^{(m)}_L[\varphi]$, thus explaining the structure of (\ref{cL}). Ultraviolet divergences are absorbed into the couplings $\Lambda$ and $\vartheta$.  Finally, we emphasise that $ \mathcal{Z}^{(m)}_{h}[\Lambda]$ is an integral over the moduli space $\mathcal{F}_h$ of the Riemann surface $\Sigma_h$, and no other geometric feature of $\tilde{g}_{ij}$, due to the absence of a conformal anomaly in the combined Liouville, ghost, and matter CFT.

\subsection{Sphere $\&$ torus partition function at large $m$}\label{sec:ST_L}

The partition function of Liouville theory $Z_L[\Lambda;S^2]$ on the round sphere of area $4\pi \upsilon$ (\ref{S2metric}) can be studied systematically in a semiclassical large $m$ expansion \cite{Zamolodvarphikov:1982vx,beatrix} (see also \cite{Mahajan:2021nsd}). Upon fixing the area of the physical metric (\ref{physmetric}) to some value $\tilde{\upsilon}$, the constant field configuration becomes a saddle point solution, permitting one to compute $Z_L[\Lambda;S^2]$ by integrating over $\tilde{\upsilon}$.
To leading order, the resulting expression is given by
\begin{equation}\label{ZLsphere}
Z^{(m)}_L[\Lambda;S^2] 
\approx \left( {\upsilon}{\Lambda_{\text{uv}}}\right)^{\frac{c_L}{6}}  (-1)^mm^{\frac{1}{2}} \, e^{-m\log m}\left(\frac{\Lambda}{\Lambda_{\text{uv}}}\right)^{m+\frac{1}{2}}~,
\end{equation}
where $\Lambda_{\text{uv}}$ is related to the UV cutoff of our theory and is independent of $\Lambda$ and $m$.
The $m$ dependent pre-factor is intricately associated to the residual $PSL(2,\mathbb{C})$ group upon fixing the Weyl gauge. Taking $\Lambda \propto \epsilon$ we observe that the non-analytic behaviour in (\ref{ZLsphere}) matches the one in (\ref{F0_path}). Thus, the identity path $\gamma_{\text{id}}^{(m)}(t)$ (\ref{gammaid}) in the matrix picture corresponds to turning on the cosmological constant $\Lambda$ which couples the identity operator of $\mathcal{M}_{2m-1,2}$ to gravity in the continuum picture. For a detailed dictionary between paths in coupling space and the couplings of primaries in $\mathcal{M}_{2m-1,2}$ we refer to \cite{penguins}.


We now consider the gravitational path integral for genus $h=1$. A torus $\mathbb{T}_\tau$ with modular parameter $\tau = \tau_1 + i \tau_2$ is locally flat and hence the Ricci scalar vanishes. On a torus we have
\begin{equation}\label{Z1}
\mathcal{Z}^{(m)}_1[\Lambda]= \int_{\mathcal{F}_1} \frac{\dd^2\tau}{\tau_2^2}\int \frac{[\mathcal{D}\varphi]}{U(1)^2}\, e^{- \frac{1}{4\pi}\,\int_{\mathbb{T}_\tau} \dd^2\sigma \, \tau_2\left(\tilde{g}^{ij}\partial_i\varphi\partial_j\varphi+ 4\pi \Lambda\, e^{2b\varphi}\right)}  {Z}_{\text{ghost}}[\Sigma_1]{Z}_{\text{CFT}}^{(m)}[\Sigma_1]~.
\end{equation}
Here we assume the line element $\dd \tilde{s}^2 = |\dd \sigma_1+ \tau \dd\sigma_2|^2$, $\sigma_i \in [0,2\pi]$ and $\mathcal{F}_1$ denotes the fundamental domain of the modular group on the upper half plane. Following the analysis in \cite{Bershadsky:1990xb,Bershadsky:1991zs} we find
\begin{equation}\label{ZT2}
\mathcal{Z}^{(m)}_{1}[\Lambda]= \frac{m-1}{48}\log\left(\text{const}\times {\Lambda}\right)~.
\end{equation}
Again, we observe the same non-analytic structure in (\ref{ZT2}) as the one stemming from the identity path in $\mathcal{F}_1^{(m)}(\epsilon)$ (\ref{F_matrix}).
{If instead of the identity we turn on one of the other $(m-2)$ primaries of $\mathcal{M}_{2m-1,2}$ we find \begin{equation}
\mathcal{Z}^{(m)}_{1}[\lambda_s]= \frac{m-1}{24m}\times \frac{m}{1+s}\log\left(\text{const}\times {\lambda_s}\right)~,\quad  s= 2,\ldots ,m-1~,
\end{equation}
where $\lambda_s$ denotes the coupling of the operator $\mathcal{O}_{1,s}$.
}


\subsection{Higher genus partition function at large $m$}\label{liouville_higherGenus}

For genus $h \ge 2$, the saddle point equations stemming from $S^{(m)}_L[\varphi]$ in (\ref{SL}) are
\begin{equation}\label{eom}
- \tilde{\nabla}^2 \varphi  - \frac{(h-1)}{\upsilon} \, Q + 4\pi \Lambda b e^{2 b \varphi} = 0~.
\end{equation}
We have chosen the fiducial metric to have constant Ricci scalar $R[\tilde{g}] = -2(h-1)/\upsilon$. The equations (\ref{eom}) admit a real constant solution
\begin{equation}
\varphi^* = \frac{1}{2b} \log \frac{(h-1)Q}{4\pi b \Lambda \upsilon}~.
\end{equation}
The on-shell action evaluated on $\varphi^*$ reads
\begin{equation}
S_L^{(m)}[\varphi^*] = \frac{Q  (h-1)}{b} \log  \frac{4\pi e \Lambda \upsilon b}{Q(h-1)} \approx -(h-1)m\log m~.
\end{equation}
We note that $S_L^{(m)}[\varphi^*]$ is independent of the moduli of $\Sigma_h$ \cite{Takhtajan:2005md}. Due to the absence of continuous conformal isometries for $\tilde{g}_{ij}$ on $\Sigma_h$  for $h \ge 2$, there is no residual gauge symmetry group whose volume we must divide by. We can compute the leading correction to the saddle point approximation by performing the Gaussian path integral over quadratic fluctuations. To Gaussian order one has for $h\ge 2$
\begin{equation}\label{ZLgen}
Z^{(m)}_L[\Lambda;\Sigma_h] \approx \left(\frac{4\pi e \Lambda \upsilon}{m(h-1)}\right)^{- (h-1)\left(m+\frac{1}{2}\right)} {\det}^{-1/2} \left(-\frac{\tilde{\nabla}^2}{\Lambda_{\text{uv}}}+\frac{2(h-1)}{\Lambda_{\text{uv}}\upsilon} \right)~,
\end{equation}
where we used that $Q/b= m+1/2$. Upon identifying $\Lambda \propto \epsilon$ we observe the same non-analytic behaviour in (\ref{ZLgen}) as the one in $\mathcal{F}_h^{(m)}(\epsilon)$ (\ref{Fhidentity}) along the identity path (\ref{gammaid}) in the matrix integral. We also note that  to leading order at large $m$ the $\upsilon$ dependence in (\ref{ZLgen}) is in accordance with the conformal anomaly as in (\ref{sewing}). 

The spectrum of the Laplacian $-\tilde{\nabla}^2$ on a Riemann surface $\Sigma_h$ depends on the moduli $(\tau_l,\bar{\tau_l})$ and has been studied, for example, in \cite{DHoker:1986eaw, Bonifacio:2020xoc, Sarnak}. Though there are few explicit results, it is known that the spectrum is positive definite. In contrast to the sphere partition function (\ref{ZLsphere}), for $h\ge 2$, there is no power law pre-factor in $m$ for $Z^{(m)}_L[\Lambda;\Sigma_h]$ for $h\ge 2$. Once a particular covariant regularisation scheme (such as the heat kernel scheme) is selected to compute the functional determinant for a given genus $h$, it must be used for all other genera. In particular, any ambiguities stemming from the choice of regularisation drop out from ratios of partition functions of different genera.

{\begin{center} \it{Sewing formula for Liouville theory?} \end{center}}
\noindent
As a final comment, we note that it may be interesting to apply the sewing formula to Liouville theory \cite{Teschner:2003at}. In this case, we must consider a complete set of states \cite{Seiberg:1990eb,Teschner:1995yf,Teschner:2001rv} which is associated to the primary operators $\mathcal{O}_\alpha = e^{2\alpha \varphi}$ with 
\begin{equation}
\alpha = \frac{Q}{2} + i P~, \quad\quad P \in \mathbb{R}^+~,
\end{equation}
and their Virasoro descendants. The conformal dimensions of $\mathcal{O}_\alpha = e^{2\alpha \varphi}$ are $\Delta_P =\bar{\Delta}_P = Q^2/4 + P^2$. We note that in the semiclassical limit, where $Q\to\infty$ with $P$ fixed, the $\Delta_P \sim c_L \sim Q^2$ are `heavy' operators, and moreover
\begin{equation}
\Delta_P - \frac{c_L}{24} \approx P^2- \frac{1}{24}~.
\end{equation}
The OPE coefficients $C^{(Q)}\left(P_{i_1},P_{i_2},P_{i_3}\right)$ of three primaries are given by the DOZZ formula \cite{Zamolodchikov:1995aa,Dorn:1994xn,Teschner:2001rv,Teschner:1995yf}
\begin{multline}\label{dozz}
C^{(Q)}\left(P_{i_1},P_{i_2},P_{i_3}\right) = \left(\pi \Lambda b^{2-2b^2} \gamma(b^2)\right)^{(Q- \sum_i \alpha_i)/b}\cr
\times \frac{\Upsilon_0 \Upsilon_b(2\alpha_1)\Upsilon_b(2\alpha_2)\Upsilon_b(2\alpha_3)}{\Upsilon_b(\sum_i \alpha_i- Q)\Upsilon_b(\alpha_1+\alpha_2-\alpha_3)\Upsilon_b(\alpha_2+\alpha_3-\alpha_1)\Upsilon_b(\alpha_3+\alpha_1-\alpha_2)}~,
\end{multline}
where 
\begin{equation}
 \log\Upsilon_b (z)= \int_0^\infty\frac{\dd t}{t}\left[\left(\frac{Q}{2}-z\right)^2e^{-t}- \frac{\sinh^2\left[\left(\frac{Q}{2}-z\right)\frac{t}{2}\right]}{\sinh \frac{tb}{2}\sinh\frac{t}{2b}}\right]~,
\end{equation}
and $\Upsilon_0 \equiv \Upsilon_b(b)$. 

It is instructive to compare the higher genus Liouville partition function (\ref{ZLgen}) to the expression obtained from applying the sewing formula (\ref{sewing}) to the Liouville CFT with central charge $c_L$ and conformal dimensions $\Delta_P= \bar{\Delta}_P$. Here we have to take into consideration that the DOZZ coefficients (\ref{dozz}) are already normalised with respect to the path integral. {\color{black} As a simple test we can compare the $\Lambda$ dependence stemming from the sewing formula to that in the semiclassical expression (\ref{ZLgen}). To do so, we note that the $C^{(Q)}\left(P_{i_1},P_{i_2},P_{i_3}\right)$  scale as $\Lambda^{-Q/2b - i(P_{i_1}+P_{i_2}+P_{i_3})/b}$. The sewing formula invokes $2h-2$ three-punctured two-spheres, so the overall  dependence on $\Lambda$ at large $Q$ is $\Lambda^{-Q(h-1)/b}$, where the $P_i$ dependent part cancels. Recalling that $Q/b=m+1/2$, this is the same $\Lambda$ dependence as the one exhibited by (\ref{ZLgen}).} Further to this, several of the building blocks of the DOZZ formula simplify in the semiclassical limit. Following \cite{Zamolodchikov:1995aa,Harlow:2011ny} we find
\begin{multline}
C^{(Q)}\left(P_{i_1},P_{i_2},P_{i_3}\right)  \approx e^{\frac{1}{3b^2}\left(\log 2- 36 \log A - 3\log b\right)} \\
\times e^{- \frac{1}{b}\left[P_{i_1}\left(2i- 2i \log (2P_{i_1})-\pi\right)+P_{i_2}\left(2i- 2i \log (2P_{i_2})-\pi\right) +P_{i_3}\left(2i- 2i \log (2P_{i_3})-\pi\right)\right]+ \ldots}\,,
\end{multline}
where we take $P_{i_k}$ of order one and have dropped the overall $\Lambda$-dependent pre-factor. Finally, we note that to first non-leading order in the semiclassical expansion, a comparison of (\ref{ZLgen}) to the sewing formula yields the schematic expression 
\begin{multline}\label{detSL}
{\det}^{-1/2} \left(-\frac{\tilde{\nabla}^2}{\Lambda_{\text{uv}}}+\frac{2(h-1)}{\Lambda_{\text{uv}}\upsilon} \right)\overset{?}{=} \lim_{Q\to\infty} \sum_{i_1,\ldots,i_{3h-3}} \int_{\mathbb{R}^+} \frac{\dd P_{i_1}}{2\pi} \ldots  \frac{\dd P_{i_{3h-3}}}{2\pi}  \\ \times \prod_{l=1}^{3h-3} \, \left( \tau_l \bar{\tau}_l \right)^{P_{i_l}^2 - \frac{1}{24}}  \times \prod_{n=1}^{2h-2} \tilde{C}^{(Q)}\left(P_{i_n},P_{j_n},P_{k_n}\right)~.
\end{multline}
In the above we have defined $\tilde{C}^{(Q)}\left(P_{i_n},P_{j_n},P_{k_n}\right)$ such that terms that go as $\sim e^{Q^2(h-1)}$  are stripped off 
such that the left and right hand side of (\ref{detSL}) do not scale with $Q$ to leading order in the large $Q$ expansion. 
 It should also be understood that the sum in (\ref{detSL}) is over Virasoro primaries and descendants, such that $\tilde{C}^{(Q)}\left(P_{i_n},P_{j_n},P_{k_n}\right)$ also includes the contribution of the Virasoro descendants.

\subsection{Higher genus ghost partition function}\label{sec:ghost}

The contribution from the ghost sector is given by the ghost determinant on $\mathcal{M}_h$. 
We define the differential operator $P$ acting on vector fields $\xi_i$ \cite{PolchinskiBook} 
\begin{equation}
P \xi_i \equiv \frac{1}{2}\left(\tilde{\nabla}_i \xi_j +\tilde{\nabla}_j \xi_i - \tilde{g}_{ij} \tilde{\nabla}_k \xi^k \right)~.
\end{equation}
The origin of $P$ comes from those diffeomorphisms that transform the metric by a traceless part. It is also convenient to introduce an object $P^T$ which transforms traceless and symmetric tensors to vector fields, namely
\begin{equation}
P^T \chi_{ij} \equiv -\tilde{\nabla}^j \chi_{ij}~, 
\end{equation}
where $\tilde{g}^{ij} \chi_{ij} =0$. We can view $P^T$ as a type of transpose of $P$. 
We note that
\begin{equation}
P^T P \xi = -\frac{1}{2} \tilde{\nabla}^j \left(\tilde{\nabla}_i \xi_j +\tilde{\nabla}_j \xi_i - \tilde{g}_{ij}\tilde{\nabla}_k \xi^k \right) = -\frac{1}{2}  \left( \tilde{\nabla}^2  + \frac{R[\tilde{g}]}{2} \right) \xi_i~,
\end{equation}
where $R[\tilde{g}]$ is the Ricci scalar of $\tilde{g}_{ij}$. Here we recall that 
\begin{equation}
[\tilde{\nabla}_i,\tilde{\nabla}_j] \xi^k = {R^k}_{lij}\xi^l~,
\end{equation}
and that in two dimensions $R_{ijkl} = R[\tilde{g}] (\tilde{g}_{ik}\tilde{g}_{jl}-\tilde{g}_{il} \tilde{g}_{jk})/2$. In terms of $P$, the ghost action is
\begin{equation}
S_{\text{ghost}} = \frac{1}{2\pi} \int_{\Sigma_h} \dd^2 x \sqrt{\tilde{g}} b_{ij} (P c)^{ij}~,
\end{equation}
where $b_{ij}$ is a traceless and symmetric Grassmann valued tensor field and $c_i$ is a Grassmann valued vector field. Thus, the ghost partition function is
\begin{equation}\label{ghost_h}
Z_{\text{ghost}}[\Sigma_h]  = \sqrt{{\det} P^T P} =  \left( \frac{\upsilon}{\upsilon_0} \right)^{\frac{26}{6}(h-1)} f_{\text{ghost}}(\tau_l,\bar{\tau}_l)~,
\end{equation}
where again we take the fiducial metric to have constant Ricci scalar $R[\tilde{g}] = -2(h-1)/\upsilon$. The $\upsilon$-dependence in (\ref{ghost_h}) is fixed by the conformal anomaly of the ghost theory which has $c_{\text{ghost}} = -26$ and $\upsilon_0$ is a reference scale. Consequently, the ghost contribution to $\mathcal{Z}^{(m)}_{\text{grav}}[\Lambda; \vartheta]$ is independent of $m$. 


%

\subsection{Comparison to matrix integrals at large $m$}\label{sec:comparison}

At this stage we can collect the various pieces and compare the leading large $m$ behaviour of matrix integral ratios of the $\mathcal{F}^{(m)}_h(\epsilon)$ such as (\ref{ratio2}) with the corresponding ratios of $\mathcal{Z}_{h}^{(m)}[\Lambda]$ from (\ref{Zgrav}).\footnote{We compare ratios of contributions stemming from different genera opposed to \cite{Goulian:1990qr,Belavin:2008kv,Belavin:2010pj} whose objects are ratios of integrated correlation functions on a fixed genus.} 
Granting the conjecture of \cite{Staudacher:1989fy,Kazakov:1989bc,Moore:1991ir} we have
\begin{equation}
\lim_{\text{d.sc.}} N^{2-2h}\mathcal{F}_h^{(m)}(\epsilon)  = \mathcal{N} \times e^{\vartheta \chi_h} \mathcal{Z}_h^{(m)}[\Lambda]~,
\end{equation}
where d.sc. denotes the double scaling employed in the derivation of (\ref{F_matrix})-(\ref{F4eps}) and $\mathcal{N}$ denotes a normalisation constant.
To leading order in the large $m$ limit, and dropping $m$-independent pre-factors, we find
\begin{eqnarray}
\mathcal{Z}_0^{(m)}[\Lambda] &\approx& m^{\frac{1}{2}} \, e^{-m \log m}\Lambda^m\times Z_{\text{CFT}}^{(m)}[S^2]~, \\
\mathcal{Z}_h^{(m)}[\Lambda] &\approx& m^{-3(h-1)} \, e^{(2\nu-1) m (h-1)} e^{m(h-1)\log m}\Lambda^{-m(h-1)}\times \left( Z_{\text{CFT}}^{(m)}[S^2] \right)^{2h-2}~, ~ h \geq 2~,
\end{eqnarray}
 where we defined $\nu$ in (\ref{opelargem}).
A particular ratio of interest that is insensitive to any overall rescaling of the path integral or matrix integral, and moreover independent of $\Lambda$, is given by
\begin{equation}\label{ratio1}
\frac{\mathcal{F}_0^{(m)}(\epsilon) \times \left( \mathcal{F}_3^{(m)}(\epsilon) \right)^2}{\left( \mathcal{F}_2^{(m)}(\epsilon) \right)^3} \overset{?}{=} \frac{\mathcal{Z}_0^{(m)}[\Lambda] \times \left( \mathcal{Z}_3^{(m)}[\Lambda] \right)^2}{\left( \mathcal{Z}_2^{(m)}[\Lambda] \right)^3}~.
\end{equation}
The matrix integral predicts that the right hand side of the above expression is independent of $m$ to leading order. From this we deduce that the sphere partition function of the $\mathcal{M}_{2m-1,2}$ minimal model in the path integral normalisation goes as 
\begin{equation}\label{zS2m}
Z_{\text{CFT}}^{(m)}[S^2] \approx  \text{const} \times \left( \frac{\upsilon}{\upsilon_0} \right)^{\frac{c_m}{6}} \times m^{\frac{5}{6}} \times e^{\frac{1}{3}\left(2\nu-1\right) \, m}~.
\end{equation} 
Equipped with $Z_{\text{CFT}}^{(m)}[S^2]$, we have sufficient information to fix the large $m$ scaling of all  $\Lambda$ independent ratios such as (\ref{ratio1}). As far as we are aware (\ref{zS2m}) has not been computed from a continuum picture. It would be very interesting to do so. Moreover, given the broader relation \cite{Douglas:1989dd} between two-dimensional gravity coupled to an arbitrary minimal model $\mathcal{M}_{p,p'}$ and matrix integrals, our general procedure seems to be an in principle novel avenue for producing a wealth of data about $Z^{(p,p')}_{\text{CFT}}[S^2]$ for more general minimal models.

It is also of interest to consider quantities that do not involve vanishing genus partition functions. For instance, we can test 
\begin{equation}
\frac{\mathcal{F}_4^{(m)}(\epsilon)}{ \mathcal{F}_2^{(m)}(\epsilon) \times \mathcal{F}_3^{(m)}(\epsilon) } \overset{?}{=} \frac{\mathcal{Z}_4^{(m)}[\Lambda] }{\mathcal{Z}_2^{(m)}[\Lambda] \times   \mathcal{Z}_3^{(m)}[\Lambda] }~.
\end{equation}
The left and right hand sides of the above quantity go as $\mathcal{O}(1)$ at large $m$. Moreover, the above quantity, though independent of $\Lambda$ and $\epsilon$ by construction, is dependent on the overall scaling of $\mathcal{F}_h^{(m)}(\epsilon)$ and $\mathcal{Z}_h^{(m)}[\Lambda]$, so it could be used to fix any such ambiguity.


One can also identify a relation between the couplings of the gravitational theory and quantities in the matrix integral. For instance, the coupling $\vartheta$ of the Euler characteristic $\chi_h$ can be identified with the logarithm of the size of the hermitian matrix $N$, that is $\vartheta   = \log N$. Moreover, the cosmological constant $\Lambda$ is proportional to $\epsilon$. A more invariant way  \cite{longsphere}  to define the couplings is through the various field redefinition invariant quantities $\mathcal{Z}_h^{(m)}[\Lambda]$. From this perspective, observing that $\mathcal{Z}_1^{(m)}[\Lambda]$ in (\ref{ZT2}) is independent of $\vartheta$ we can use it to define the cosmological constant $\Lambda$. Once $\Lambda$ is defined through $\mathcal{Z}_1^{(m)}[\Lambda]$, we can proceed to define $\vartheta$ through one with $h \neq 1$ terms in $\mathcal{Z}^{(m)}_{\text{grav}}[\Lambda; \vartheta]$. In such a way, one may hope to eliminate any remaining ambiguities allowing for a precise comparison between the infinite functions of $m$, i.e. $\mathcal{Z}_h^{(m)}[\Lambda]$ and $\mathcal{F}_h^{(m)}(\epsilon)$ with $h=0,1,...$, stemming from the matrix integral and continuum pictures.

%

\section{Remarks on a two-dimensional de Sitter universe}\label{sec:dS}

In this section we comment on the potential relation of our two-dimensional gravitational theories to thermodynamic considerations of the de Sitter horizon in two spacetime dimensions.\footnote{Recent activity on microscopic models for the static patch de Sitter horizon include \cite{Shyam:2021ciy,Coleman:2021nor}, where the microstates of the dS$_3$ are argued to be obtained by a $T\bar{T}$ reassembling  of the BTZ microstates, and \cite{Anninos:2017hhn,Anninos:2018svg,Anninos:2020cwo} where the microstates of dS$_2$ are argued to be obtained by reassembling those of an AdS$_2$ black hole in SYK type theories. The interior of the de Sitter horizon was explored in \cite{Damian_Shira,Susskind:2021esx,Shaghoulian:2021cef}.}

\subsection{Entropic hints at large $m$}


To obtain a saddle point expansion about the round two-sphere geometry on an $S^2$ topology in the limit of large and negative $c_m$ (the large $m$ limit) one must fix the area $\upsilon$ of the physical metric \cite{Zamolodvarphikov:1982vx,Seiberg:1990eb,beatrix}.  
Thus, rather than $\mathcal{Z}^{(m)}_{0}[\Lambda]$ in (\ref{Zgrav0}), it is
\begin{equation}\label{fixedA1}
\tilde{\mathcal{Z}}^{(m)}_{0}[\Lambda;\upsilon] = \int [\mathcal{D} g] e^{-\Lambda \int_{S^2} \dd^2 x\sqrt{g}} \, Z_{\text{CFT}}^{(m)}[g_{ij} ;\Sigma_h] \times \delta\left(\int_{S^2} \dd^2 x \sqrt{g} - 4\pi\upsilon \right)~,
\end{equation}
that exhibits a semiclassical saddle given by the round two-sphere at large $m$. 
At fixed area $\upsilon$ the gravitational path integral (\ref{fixedA1}) yields 
\begin{equation}\label{Zgrav_disc2}
\log {e^{2\vartheta}} \mathcal{\tilde{Z}}^{(m)}_{0}[\upsilon] =  2\vartheta - \left(\frac{24}{\left(\sqrt{1-c_m}- \sqrt{25-c_m}\right)^2}+2\right) \log\frac{\upsilon}{\upsilon_0} + \tilde{f}^{(0)}(c_m)~,
\end{equation}
where $\tilde{f}^{(0)}(c_m)$ captures higher loop contributions \cite{beatrix,timelike}. The coefficient of the logarithmic term \cite{KPZ,Brezin:1989db} admits a semiclassical $c_m \rightarrow -\infty$ expansion
\begin{equation}\label{logcoeff}
 {-} \left(   \frac{24}{\left(\sqrt{1-c_m}- \sqrt{25-c_m}\right)^2}+2 \right) \approx \left(\frac{c_{m}-26+1}{6}-\frac{6}{c_m}+ \ldots \right)~.
 \end{equation}
The leading term of (\ref{logcoeff}) is identical to that of a two-dimensional CFT on a rigid round two-sphere of area $\upsilon$ of central charge $c_m-26+1$. Perhaps, in the spirit of the Gibbons-Hawking conjecture, we should interpret this as the entanglement entropy from the matter-ghost-Liouville sector \cite{Calabrese:2004eu,Holzhey:1994we,Casini:2011kv} with the subleading corrections coming from dynamical gravity \cite{musings,penguins}. The leading term $2\vartheta= 2\log N$ can be viewed as a two-dimensional version of the tree-level horizon entropy. 

We can also transform from (\ref{fixedA1}) to (\ref{Zgrav0}) using an inverse Laplace transform, recovering
\begin{equation}
\log {e^{2\vartheta}}  \mathcal{Z}_{0}^{(m)}[\Lambda]= 2\vartheta + \left(\frac{24}{\left(\sqrt{1-c_m}- \sqrt{25-c_m}\right)^2}+1\right)\log \frac{\Lambda}{\Lambda_{\text{uv}}}+ f^{(0)}(c_m)~.
\end{equation}
It is worth recalling here the structure of $\mathcal{Z}_{0}[\Lambda]$ in $(d+1)$ dimensions with $d\geq 2$ \cite{longsphere} on a $(d+1)$-dimensional sphere expanded with respect to the tree-level de Sitter entropy $\mathcal{S}_0 \gg 1$:
\begin{equation}
\log \mathcal{Z}_{0}[\Lambda] = \mathcal{S}_0 + a_0 \log \mathcal{S}_0 -  a_1 \log \frac{\Lambda}{\Lambda_{\text{uv}}} + a_2 + \ldots~.
\end{equation}
The coefficient $a_0$ is a group theoretic factor stemming from the residual diffeomorphisms, while $a_1$ is a more standard coefficient of the logarithmic divergence at one-loop. For pure gravity in four-dimensions we have $\mathcal{S}_0 = 3\pi/(\Lambda G)$, $a_0 = - 5$, and $a_1 = -571/90$, while $a_2$ is a computable constant that depends on the regularisation scheme. 

Such entropic hints prompt us to consider the possibility of a Lorentzian picture.

\subsection{Toward a Lorentzian picture}\label{Lorentzian}

We would like to end our discussion by emphasising a crucial challenge in relating our computations to those of a semiclassical Lorentzian de Sitter world. In Euclidean signature, the path-integral exhibiting a saddle point geometry given by the round two-sphere requires that we fix the physical area of the metric (\ref{fixedA1}) \cite{Zamolodvarphikov:1982vx,beatrix}.  This is in stark contrast to the higher dimensional case, or the case where $c_m$ is large and positive \cite{timelike,Polchinski:1989fn}, for which the round sphere is a saddle of the original path integral. Relatedly, (\ref{fixedA1})  is an inherently Euclidean expression --  the round two-sphere saddle of (\ref{fixedA1}) has seemingly little Lorentzian meaning. 

Perhaps a way to circumvent this is to introduce a Lagrange multiplier for the $\delta$-function such that
\begin{equation}\label{delta}
\delta\left( \int_{S^2} \dd^2x\sqrt{g} - 4\pi \upsilon \right) = \int_{\mathbb{R}} \frac{\dd\alpha}{2\pi} \exp {i \alpha \left( \int_{S^2} \dd^2x\sqrt{g} - 4\pi\upsilon \right)}~.
\end{equation}
From this perspective, the cosmological constant $\Lambda_\alpha \equiv \Lambda - i \alpha$ and we integrate\footnote{Integrating over couplings is something that is occasionally done in physics, particularly for systems with disorder/random impurities. More recently, the idea of averaging over couplings has appeared in the context of black holes and lower dimensional theories of gravity \cite{Kitaev:2017awl,Anninos:2013nra,Saad:2019lba}. The current context is perhaps closer in spirit to the ideas in \cite{Coleman:1988tj,Klebanov:1988eh}.}  $\alpha$ against the distribution function $f_\upsilon(\alpha) = e^{-4\pi i \alpha \upsilon}$ where $\upsilon$ is viewed as a parameter of the distribution function. 
The large $m$ saddle point solutions for $\alpha$ and $\varphi$ are
\begin{equation}
\alpha_* = - i \frac{Q}{4\pi b {\color{black} \upsilon}}~, \quad\quad \varphi_* = 0~.
\end{equation}
Fluctuations $(\delta\alpha,\delta\varphi)$ about the saddle point solution $(\alpha_*, \varphi_*)$ are governed by the small coupling $1/m$. Interestingly, although the Euclidean theory is not reflection positive due to a complex $\Lambda_\alpha$, both the saddle $\alpha_*$ and the fluctuations $\delta\alpha$ live along the pure imaginary axis \cite{beatrix}, thereby restoring reflection positivity. In view of (\ref{delta}), we might propose the following Lorentzian action in the Weyl gauge (\ref{physmetric}):
\begin{equation}
S^{(\alpha)}_L[\varphi] = \frac{1}{4\pi}\int_{\mathbb{R}} \dd T \dd\phi \cosh T \left((\partial_T \varphi)^2 -\frac{(\partial_\phi \varphi)^2}{\cosh^2 T}  - 2 Q   \varphi - 4\pi \upsilon \Lambda_\alpha  e^{2b\varphi}\right)~,
\end{equation}
where we take $\theta \to i T+\pi/2$ in (\ref{S2metric}). The theory is prepared in the Hartle-Hawking \cite{HH} state $|\text{HH}_\upsilon\rangle$,  and one computes expectation values $\langle \text{HH}_\upsilon | \hat{O} | \text{HH}_\upsilon \rangle$ for some observable $\hat{O}$. Finally, we integrate $\alpha$ over the distribution $f_\upsilon(\alpha)$. If this is a reasonable Lorentzian picture, and given the conjectural relation of (\ref{fixedA1}) to the matrix integral (\ref{MI}), the setup under consideration offers a concrete approach toward a microscopic understanding of a dS$_2$ universe. Time will tell. 

\section*{Acknowledgements}
It is a pleasure to acknowledge Simon Caron-Huot, Frederik Denef, Elias Kiritsis, Alex Maloney, John Stout and Gerard Watts for useful discussions. 
D.A. is funded by the Royal Society under the grant The Atoms of a deSitter Universe. B.M. is supported in part by the Simons Foundation Grant No. 385602 and the Natural Sciences and Engineering Research Council of Canada (NSERC), funding reference number SAPIN/00047-2020.

\appendix{

\section{Paths in coupling space}\label{app:paths}
In this appendix we review the main concepts of \cite{musings,penguins,Ginsparg:1993is,DiFrancesco:1993cyw} formulated in terms of paths (\ref{paths}) in coupling space. Employing large $N$ techniques, the planar limit of the matrix integral (\ref{MI}) is captured by
\begin{equation}\label{eq:actioneig}
S[\rho^{(m)}_{\mathrm{ext}}(\lambda,\boldsymbol{\alpha})] =\frac{1}{2}\int_{-a}^a\dd \lambda\rho^{(m)}_{\mathrm{ext}}(\lambda,\boldsymbol{\alpha})V_m(\lambda,\boldsymbol{\alpha})- 2\int_0^a\dd \lambda\rho^{(m)}_{\mathrm{ext}}(\lambda,\boldsymbol{\alpha})\log(\lambda)~,
\end{equation}
where $\lambda$ are the eigenvalues of the Hermitian matrix $M$ which we assume are distributed in the interval $[-a,a]$ along the real axis, and $\rho^{(m)}_{\mathrm{ext}}(\lambda,\boldsymbol{\alpha})$ is the eigenvalue distribution obtained from the large $N$ saddle point approximation
\begin{align}\label{eq:eigdistrV}
\rho^{(m)}_{\mathrm{ext}}(\lambda,\boldsymbol{\alpha})&= \frac{1}{\pi}\sum_{n=1}^m \frac{\alpha_n}{B(n,1/2)}\lambda^{2n-2}\,_2 F_1\left(\frac{1}{2},1-n; \frac{3}{2}; 1- \frac{u}{\lambda^2}\right)\sqrt{u-\lambda^2}~.
\end{align}
Demanding that this distribution is normalised we obtain for the $m^{\text{th}}$ multicritical matrix integral an $m^{\text{th}}$ order polynomial constraint
\begin{equation}\label{eq:BC}
0=  1- \sum_{n=1}^m\frac{\alpha_n u^n}{2n B(n,1/2)}~, \quad \alpha_1 \equiv 1~,\quad\quad u\equiv a^2~.
\end{equation}
This polynomial has in general $m$ distinct solutions, however choosing the exponents in (\ref{paths}) appropriately one can collapse several of these solutions to have higher multiplicity. For the identity path we have one second order solution and $(m-2)$ distinct other solutions. On the other extreme along the `minimal' path we have one $m^{\text{th}}$ order solution. 
\newline\newline
\textbf{Example $m=3$}: In this case the normalisation condition (\ref{eq:BC}) is a third order polynomial 
\begin{equation}\label{norm3}
1- \frac{1}{4}u - \frac{3}{16}\alpha_2 u^2- \frac{5}{32}\alpha_3 u^3=0~.
\end{equation}
The critical point lies at $\alpha_{2,c}^{(3)}= -1/9$ and $\alpha_{3,c}^{(3)}= 1/270$. 
We distinguish the two paths 
\begin{equation}
\gamma_{\text{min}}^{(3)}(t)= \begin{pmatrix} \alpha_{2,c}^{(3)}t \\ \alpha_{3,c}^{(3)}t^2 \end{pmatrix}~,\quad \quad \gamma_{\text{id}}^{(3)}(t)= \begin{pmatrix} \alpha_{2,c}^{(3)}t^{1/3} \\ \alpha_{3,c}^{(3)}t \end{pmatrix}~,\quad t\in [0,1]~.
\end{equation}
As we approach the multicritical point along $\gamma_{\text{min}}^{(3)}(t)$ the normalisation condition (\ref{norm3}) exhibits a third order zero for $u=12$. Along $\gamma_{\text{id}}^{(3)}(t)$ we find a second order zero. 
\newline\newline
It is straightforward to generalise the above example to arbitrary $m$ using the expressions in \cite{penguins}. For a  small deviation $t=1-\epsilon$, $0< \epsilon \ll 1$ we can explore the non-analyticity of $\mathcal{F}_m^{(0)}$. We find the leading planar non-analyticity  
\begin{equation}
\mathcal{F}_{m}^{(0)} (\epsilon) = \frac{4}{(2m-3)(2m-1)(2m+1)}\epsilon^{m+\frac{1}{2}}~, 
\end{equation}
along the identity path and
\begin{equation}
\mathcal{F}_{m}^{(0)} (\epsilon)  = \frac{m^2}{(m+1)(2m+1)}\epsilon^{2+\frac{1}{m}}~,
\end{equation}
along the minimal path. 
Both cases exactly reproduce the results obtained using the string equation (\ref{F_matrix}). 

\section{String equation}\label{app:SE}
In this appendix we provide some details leading to the higher genus contributions of the matrix free energy (\ref{F_matrix}) following \cite{Belavin:2008kv,Belavin:2010pj}. Whereas genus zero is efficiently computed using a large $N$ saddle point approximation, the method of orthogonal polynomials provides a systematic way to compute non-planar contributions. Moreover, orthogonal polynomials lead to the string equation \cite{Douglas:1989ve,Gross:1989vs,Brezin:1990rb} which captures the full genus expansion in terms of a differential equation. For multicritical matrix integrals the string equation  takes the form \cite{Belavin:2008kv,Belavin:2010pj}}
\begin{equation}
\sum_{k=1}^{m}t_{m-k-2}\mathcal{R}_k[u(z)] + z = 0~,\quad t_{-2} \equiv - 1~,\quad  t_{-1} \equiv 0~, \quad t_{0} \equiv  \epsilon~.
\end{equation} 
where $\mathcal{R}[u(z)]$ are known as the Gelfand-Dikii polynomials, defined recursively 
\begin{equation}\label{string_equation}
\mathcal{R}_0[u(z)]= \frac{1}{2}~, \quad \mathcal{R}_{k+1}'[u(z)]= \frac{\varepsilon^2}{4}\mathcal{R}_k'''[u(z)]-u(z)\mathcal{R}_k'[u(z)]-\frac{1}{2}u'(z)\mathcal{R}_k[u(z)]~,
\end{equation}
and we have 
\begin{equation}\label{string_eq}
\frac{(-1)^{k} 2^{2k}}{\binom{2k-1}{k-1}}\mathcal{R}_k[u(z)]=u(z)^k+ \mathcal{O}(\varepsilon)~.
\end{equation}
The string equation and the free energy (\ref{planar}) are related by \cite{Douglas:1989ve,Gross:1989vs,Brezin:1990rb}
\begin{equation}\label{string_equation_app}
u(z) = \frac{\dd^2}{\dd z^2}\mathcal{F}^{(m)}~.
\end{equation}
For the $m^{\text{th}}$ multicritical matrix integral, we have
\begin{equation}
0= \mathcal{P}_m(u(z)) +\sum_{n\geq 1} \varepsilon^{2n} \mathcal{R}_{\varepsilon^{2n}}[u(z)]~,
\end{equation}
where $\mathcal{R}_{\varepsilon^{2n}}[u(z)]$ captures the order $\mathcal{O}(\varepsilon^{2n})$ contribution of (\ref{string_equation}). 
Rescaling the coefficients $t_{m-k-2}\rightarrow -t_{m-k-2}  \times (-1)^k 2^k /  \binom{2k-1}{k-1}$ we can introduce the polynomial 
\begin{equation}
\mathcal{P}_m(u(z)) = u(z)^m - \epsilon \, u(z)^{m-2}  -   t_{m-3} u(z)^{m-3} - \ldots - z~, 
\end{equation}
and obtain
\allowdisplaybreaks
\begin{align}
&\mathcal{R}_{\varepsilon^2}[u(z)]=-\frac{1}{12}\PP^{(3)}(u(z))u'(z)^2- \frac{1}{6} {\PP}''(u(z))u''(z)~,\cr
&\mathcal{R}_{\varepsilon^4}[u(z)]=  \frac{1}{288}\PP^{(6)}(u(z))u'(z)^4+\frac{1}{60}\PP^{(3)}(u(z))u^{(4)}(z)+ \frac{1}{40}\PP^{(4)}(u(z))u^{(2)}(z)^2\cr
&+ \frac{1}{30}\PP^{(4)}(u(z))u'(z)u^{(3)}(z)+ \frac{11}{360}\PP^{(5)}(u(z))u'(z)^2u{''}(z)~,\cr
&\mathcal{R}_{\varepsilon^6}[u(z)]=- \frac{1}{10368}\PP^{(9)}(u(z))u'(z)^{6} - \frac{17}{8640}\PP^{(8)}(u(z))u'(z)^{4}u''(z)\cr
&- \frac{83 }{10080}\PP^{(7)}(u(z))u'(z)^2 u''(z)^2- \frac{61}{15120}\PP^{(6)}(u(z))u''(z)^3- \frac{1}{252}\PP^{(7)}(u(z))u'(z)^3u'''(z)\cr
&- \frac{43}{2520}\PP^{(6)}(u(z))u'(z)u''(z)u'''(z)- \frac{23}{5040}\PP^{(5)}(u(z))u'''(z)^2- \frac{5}{1008}\PP^{(6)}(u(z))u'(z)^2u^{(4)}(z)\cr
&- \frac{19}{2520}\PP^{(5)}(u(z))u''(z)u^{(4)}(z)-\frac{1}{280}\PP^{(5)}(u(z))u'(z)u^{(5)}(z)- \frac{1}{840}\PP^{(4)}(u(z))u^{(6)}(z).
\end{align}
For notational convenience we suppressed the subscript $m$ in $\mathcal{P}_m(u(z)) $.
Making the ansatz 
\begin{equation}
u(z)= u_0(z)+ \varepsilon^2 u_1(z) + \varepsilon^4 u_2(z)+ \ldots~,
\end{equation}
and again grouping expressions in orders of $\varepsilon$ we obtain recursive equations for $u_n(z)$, with $n\geq 1$, in terms of $u_0(z)$. For example at order $\varepsilon^2$  we obtain
\begin{align}
u_1(z)  = \frac{1}{12}\frac{\PP_m'''(u_0)}{\PP_m'(u_0)} u_0'(z)^2+ \frac{1}{6}\frac{\PP_m''(u_0)}{\PP_m'(u_0)}u_0''(z)~.
\end{align}
Analogously, we obtain an equation for $u_n(z)$, for $n\geq 2$. Now we use 
\begin{equation}
 \PP_m(u_0(z);z) =0~,
\end{equation}
which allows us to relate $z$-derivatives of $u_0(z)$ with $u_0$-derivatives of $\mathcal{P}_m(u_0)$. The semicolon in $\PP_m(u_0(z);z)$ indicates that there is implicit and explicit dependence on $z$ in $ \PP_m(u_0(z))$. As an example we find
\begin{equation}
u_0'(z) = \frac{1}{\mathcal{P}_m'(u_0)}~,\quad u_0''(z)= - \frac{\mathcal{P}_m''(u_0)}{\mathcal{P}_m'(u_0)^3}~,~\ldots~.
\end{equation}
Solving the order $\mathcal{O}(\varepsilon^{2n})$ equation for $u_n(z)$ we obtain $\mathcal{F}^{(m)}_h$  upon integrating (\ref{string_equation_app})
\begin{equation}
\mathcal{F}^{(m)}_h= -\int_0^{u^*(z)} \dd u_0 \mathcal{P}_m(u_0;z)\mathcal{P}_m'(u_0) u_h(u_0)~,
\end{equation}
where $u^*$ is the solution of $\PP_m (u^*)=0$, and $u_h$ is to be understood as a function of $u_0$. Following the steps performed for $h=0,1,2,$ and $h=3$ we obtain
\begin{align}
&\mathcal{F}^{(m)}_4= -\frac{1}{87091200 \PP'(u)^{15}}\Big(
145349680 \PP''(u)^9-532202720 \PP'(u)\PP''(u)^7 \PP'''(u)\cr
&+ 1400 \PP'(u)^2 \PP''(u)^5 \left(437079 \PP''(u) +113176 \PP''(u)\PP^{(4)}(u)\right)-560 \PP'(u)^3 \PP''(u)^3 \bigg(503106 \PP^{(4)}(u) \PP^{(3)}(u) \PP''(u)\cr
&+63987 \PP^{(5)}(u) \PP''(u)^2+432010 \PP^{(3)}(u)^3\bigg)+
35 \PP'(u)^4 \PP''(u) \bigg(3122040 \PP^{(4)}(u) \PP^{(3)}(u)^2 \PP''(u)\cr
&+1323168 \PP^{(5)}(u)\PP^{(3)}(u) \PP''(u)^2
+72 \PP''(u)^2 \left(2471 \PP^{(6)}(u) \PP''(u)+11218\PP^{(4)}(u)^2\right)+671165 \PP^{(3)}(u)^4\bigg)\cr
&-28 \PP'(u)^5 \bigg(352704 \PP^{(5)}(u) \PP^{(3)}(u)^2 \PP''(u)+6 \PP^{(3)}(u) \PP''(u) \left(31588
   \PP^{(6)}(u) \PP''(u)+71685 \PP^{(4)}(u)^2\right)\cr
&+6 \PP''(u)^2 \left(4907 \PP^{(7)}(u)\PP''(u)+45582 \PP^{(4)}(u) \PP^{(5)}(u)\right)+185251 \PP^{(4)}(u) \PP^{(3)}(u)^3\bigg)\cr
 &+ \PP'(u)^6 \bigg(120 \PP^{(4)}(u) \left(5405 \PP^{(6)}(u) \PP''(u)+10066 \PP^{(3)}(u)
   \PP^{(5)}(u)\right)
 +24 \PP''(u) \Big(3339 \PP^{(8)}(u) \PP''(u)\cr
 &+15990 \PP^{(5)}(u)^2+16243\PP^{(3)}(u) \PP^{(7)}(u)\Big)
 +245000 \PP^{(4)}(u)^3+419370 \PP^{(3)}(u)^2 \PP^{(6)}(u)\bigg)+ 175 \PP^{(10)}(u) \PP'(u)^8\cr
 &-5 \PP'(u)^7 \left(1043 \PP^{(9)}(u) \PP''(u)+7305 \PP^{(5)}(u)
   \PP^{(6)}(u)+5343 \PP^{(4)}(u) \PP^{(7)}(u)+2841 \PP^{(3)}(u) \PP^{(8)}(u)\right)
\Big)~,\cr
\end{align}
which we evaluate at $u^*$.
To conclude, we emphasise that it is straightforward to continue along this line and calculate the free energy at arbitrary genus $h$. 
\section{General OPE coefficients at large $m$}\label{minimal}
The main objective of this appendix is to provide some details toward the derivation of (\ref{opegen}), in particular of the $m$ independent factor $\upsilon_{r_1,r_2}^{r_3}$. To do so we make use of the Euler-MacLaurin formula \cite{wiki}. For $p \geq 0$ a positive integer and a function $f(x) \in \mathcal{C}^p[m,n]$ the Euler-MacLaurin formula states 
\begin{equation}\label{EM}
\sum_{i=m}^n f(i)= \int_m^n \dd xf(x) + \frac{f(n)+f(m)}{2}+ \sum_{k=1}^{\lfloor \frac{p}{2}\rfloor} \frac{B_{2k}}{(2k)!} f^{(2k-1)}(x)\big|_{x=n}^{m}+ R_p~,
\end{equation}
where $R_p$ is a rest term which can be estimated as 
\begin{equation}\label{error_EM}
|R_p| \leq \frac{2\zeta(p)}{(2\pi)^p}\int_m^n  \dd x |f^{(p)}(x)|~.
\end{equation}
To obtain the large $m$ approximation of the OPE coefficients we apply the general Euler-Maclaurin formula (\ref{EM}) to $\log \mathcal{C}_{m-r_1,m-r_2}^{m-r_3}$ (\ref{OPE_DF}) for $p=1$. Before doing so however we realise that whereas the final expression (\ref{OPE_DF}) is permutation invariant (up to at most a sign) in its indices $\{s_1,s_2,s_3\}$ this invariance is not manifest in the individual pieces in (\ref{OPE_DF}) and a naive application of the Euler-MacLaurin formula therefore yields an expression of $\upsilon_{r_1,r_2}^{r_3}$ lacking to exhibit this invariance. 
Provided
\begin{equation}
\text{max}\{\frac{-s_1+s_2+s_3+1}{2},\frac{s_1-s_2+s_3+1}{2},\frac{s_1+s_2-s_3+1}{2}\} <\text{min}\{s_1,s_2,s_3\}~
\end{equation}
we can rewrite the OPE coefficients (\ref{OPE_DF}) as 
\begin{equation}
C_{s_1,s_2}^{s_3}= \begin{cases} (-1)^{s_3+1}(a_{s_1}a_{s_2}a_{s_3})^{1/2}\times \left( \frac{1}{a_{s_3}}\times D_{s_1,s_2}^{s_3}\right)~,\quad s_3<m~,\\
 (-1)^{s_3}(a_{s_1}a_{s_2}a_{s_3})^{1/2}\times \left( \frac{1}{a_{s_3}}\times D_{s_1,s_2}^{s_3}\right)~,\quad s_3\geq m~,
\end{cases}~
\end{equation}
where 
\begin{multline}
 \frac{1}{a_{s_3}}\times D_{s_1,s_2}^{s_3}=\prod_{i=1}^{\frac{1}{2}\lfloor s_1+s_2+s_3-3\rfloor}{\gamma(-1+(1+i)\rho)}\cr
 \times \prod_{i=\frac{1}{2}(-s_1+s_2+s_3)}^{s_1-1}\frac{1}{\gamma(i\rho)}\times  \prod_{i=\frac{1}{2}(-s_2+s_1+s_3)}^{s_2-1}\frac{1}{\gamma(i\rho)}\times \prod_{i=\frac{1}{2}(-s_3+s_1+s_2)}^{s_3-1}\frac{1}{\gamma(i\rho)}~,
\end{multline}
It is now straightforward to apply (\ref{EM}) leading to 
\begin{equation}
\lim_{m\rightarrow \infty} C_{m-r_1,m-r_2}^{m-r_3} \approx \upsilon_{r_1,r_2}^{r_3} m^{-3/2}e^{\alpha m}~,
\end{equation}
where $\alpha\equiv 12\log A-1-(\log 2)/3$ and 
\begin{align}
&|\upsilon_{r_1,r_2}^{r_3}|\cr
&= \frac{e^{-\frac{3}{2}-\frac{25}{12}\log 2}}{A^6}\times \frac{(1+2r_1)^{r_1/2}(1+2r_2)^{r_2/2}(1+2r_3)^{r_3/2}}{(1-2r_1)^{r_1/2}(1-2r_2)^{r_2/2}(1-2r_3)^{r_3/2}} \frac{{(1-2r_1)^{1/2}(1-2r_2)^{1/2}(1-2r_3)^{1/2}}}{2^{r_1+r_2+r_3}}\cr
&\times e^{\frac{1}{12(1+2r_1)}+ \frac{1}{12(1+2r_2)}+\frac{1}{12(1+2r_3)}}~.
\end{align}
The error between the sum and the integral is at most (\ref{error_EM}) 
\begin{equation}
|R_2| \leq \frac{1}{12}e^{\frac{5}{4}+\frac{1}{1-2r_1}+ \frac{1}{1+2r_1}+\frac{1}{1-2r_2}+ \frac{1}{1+2r_2}+\frac{1}{1-2r_3}+ \frac{1}{1+2r_3}}+ \mathcal{O}(m^{-2})~,
\end{equation}
which is of the order as $\upsilon_{r_1,r_2}^{r_3}$.

}

 \begingroup


\addcontentsline{toc}{section}{References}
\section*{References}
\begin{thebibliography}{99}




\bibitem{Gibbons:1976ue}
G.~W.~Gibbons and S.~W.~Hawking,
``Action Integrals and Partition Functions in Quantum Gravity,''
Phys. Rev. D \textbf{15}, 2752-2756 (1977)
doi:10.1103/PhysRevD.15.2752
  
\bibitem{Gibbons:1977mu}
G.~W.~Gibbons and S.~W.~Hawking,
``Cosmological Event Horizons, Thermodynamics, and Particle Creation,''
Phys. Rev. D \textbf{15}, 2738-2751 (1977)
doi:10.1103/PhysRevD.15.2738

\bibitem{longsphere}
D.~Anninos, F.~Denef, Y.~T.~A.~Law and Z.~Sun,
``Quantum de Sitter horizon entropy from quasicanonical bulk, edge, sphere and topological string partition functions,''
[arXiv:2009.12464 [hep-th]].

\bibitem{Law:2020cpj}
Y.~T.~A.~Law,
``A Compendium of Sphere Path Integrals,''
[arXiv:2012.06345 [hep-th]].

\bibitem{David:2021wrw}
J.~R.~David and J.~Mukherjee,
``Partition functions of $p$-forms from Harish-Chandra characters,''
doi:10.1007/JHEP09(2021)094
[arXiv:2105.03662 [hep-th]].

\bibitem{Anninos:2021ihe}
D.~Anninos and E.~Harris,
``Three-dimensional de Sitter horizon thermodynamics,''
JHEP \textbf{10}, 091 (2021)
doi:10.1007/JHEP10(2021)091
[arXiv:2106.13832 [hep-th]].

\bibitem{Hikida:2021ese}
Y.~Hikida, T.~Nishioka, T.~Takayanagi and Y.~Taki,
``Holography in de Sitter Space via Chern-Simons Gauge Theory,''
[arXiv:2110.03197 [hep-th]].

\bibitem{timelike}
D.~Anninos, T.~Bautista and B.~M\"uhlmann,
``The two-sphere partition function in two-dimensional quantum gravity,''
JHEP \textbf{09}, 116 (2021)
doi:10.1007/JHEP09(2021)116
[arXiv:2106.01665 [hep-th]].

\bibitem{musings}
D.~Anninos and B.~M\"uhlmann,
``Notes on matrix models (matrix musings),''
J. Stat. Mech. \textbf{2008}, 083109 (2020)
doi:10.1088/1742-5468/aba499
[arXiv:2004.01171 [hep-th]].

\bibitem{penguins}
D.~Anninos and B.~M\"uhlmann,
``Matrix integrals \& finite holography,''
JHEP \textbf{06}, 120 (2021)
doi:10.1007/JHEP06(2021)120
[arXiv:2012.05224 [hep-th]].

\bibitem{beatrix}
B.~M\"uhlmann,
``The two-sphere partition function in two-dimensional quantum gravity at fixed area,''
doi:10.1007/JHEP09(2021)189
[arXiv:2106.04532 [hep-th]].





\bibitem{Douglas:1989ve} 
  M.~R.~Douglas and S.~H.~Shenker,
  ``Strings in Less Than One-Dimension,''
  Nucl.\ Phys.\ B {\bf 335}, 635 (1990).
  doi:10.1016/0550-3213(90)90522-F
 
\bibitem{Gross:1989vs} 
  D.~J.~Gross and A.~A.~Migdal,
  ``Nonperturbative Two-Dimensional Quantum Gravity,''
  Phys.\ Rev.\ Lett.\  {\bf 64}, 127 (1990).
  doi:10.1103/PhysRevLett.64.127
       
\bibitem{Brezin:1990rb} 
  E.~Brezin and V.~A.~Kazakov,
  ``Exactly Solvable Field Theories of Closed Strings,''
  Phys.\ Lett.\ B {\bf 236}, 144 (1990).
  doi:10.1016/0370-2693(90)90818-Q



\bibitem{Kazakov:1989bc}
V.~Kazakov,
``The Appearance of Matter Fields from Quantum Fluctuations of 2D Gravity,''
Mod. Phys. Lett. A \textbf{4}, 2125 (1989)
doi:10.1142/S0217732389002392

%
\bibitem{Staudacher:1989fy}
M.~Staudacher,
``The Yang-lee Edge Singularity on a Dynamical Planar Random Surface,''
Nucl. Phys. B \textbf{336}, 349 (1990)
doi:10.1016/0550-3213(90)90432-D


\bibitem{Moore:1991ir}
G.~W.~Moore, N.~Seiberg and M.~Staudacher,
``From loops to states in 2-D quantum gravity,''
Nucl. Phys. B \textbf{362}, 665-709 (1991)
doi:10.1016/0550-3213(91)90548-C


\bibitem{Pestun:2016zxk}
V.~Pestun, M.~Zabzine, F.~Benini, T.~Dimofte, T.~T.~Dumitrescu, K.~Hosomichi, S.~Kim, K.~Lee, B.~Le Floch and M.~Marino, \textit{et al.}
``Localization techniques in quantum field theories,''
J. Phys. A \textbf{50}, no.44, 440301 (2017)
doi:10.1088/1751-8121/aa63c1
[arXiv:1608.02952 [hep-th]].

\bibitem{Marino:2002fk}
M.~Marino,
``Chern-Simons theory, matrix integrals, and perturbative three manifold invariants,''
Commun. Math. Phys. \textbf{253}, 25-49 (2004)
doi:10.1007/s00220-004-1194-4
[arXiv:hep-th/0207096 [hep-th]].

\bibitem{Tierz:2002jj}
M.~Tierz,
``Soft matrix models and Chern-Simons partition functions,''
Mod. Phys. Lett. A \textbf{19}, 1365-1378 (2004)
doi:10.1142/S0217732304014100
[arXiv:hep-th/0212128 [hep-th]].

\bibitem{Anninos:2016klf}
D.~Anninos and G.~A.~Silva,
``Solvable Quantum Grassmann Matrices,''
J. Stat. Mech. \textbf{1704}, no.4, 043102 (2017)
doi:10.1088/1742-5468/aa668f
[arXiv:1612.03795 [hep-th]].



\bibitem{Zamolodvarphikov:1982vx} 
  A.~B.~Zamolodchikov,
  ``On The Entropy Of Random Surfaces,''
  Phys.\ Lett.\  {\bf 117B}, 87 (1982).
  doi:10.1016/0370-2693(82)90879-6

\bibitem{Takhtajan:2005md}
L.~A.~Takhtajan and L.~P.~Teo,
``Quantum Liouville theory in the background field formalism. I. Compact Riemann surfaces,''
Commun. Math. Phys. \textbf{268}, 135-197 (2006)
doi:10.1007/s00220-006-0091-4
[arXiv:hep-th/0508188 [hep-th]].


\bibitem{Ginsparg:1982rs}
P.~H.~Ginsparg and M.~J.~Perry,
``Semiclassical Perdurance of de Sitter Space,''
Nucl. Phys. B \textbf{222}, 245-268 (1983)
doi:10.1016/0550-3213(83)90636-3
\bibitem{Polchinski:1994fq}
J.~Polchinski,
``Combinatorics of boundaries in string theory,''
Phys. Rev. D \textbf{50}, R6041-R6045 (1994)
doi:10.1103/PhysRevD.50.R6041
[arXiv:hep-th/9407031 [hep-th]].

\bibitem{Ginsparg:1991ws} 
  P.~H.~Ginsparg and J.~Zinn-Justin,
  ``Large order behavior of nonperturbative gravity,''
  Phys.\ Lett.\ B {\bf 255}, 189 (1991).
  doi:10.1016/0370-2693(91)90234-H
 
\bibitem{Eynard:1992sg} 
  B.~Eynard and J.~Zinn-Justin,
  ``Large order behavior of 2-D gravity coupled to d $<$ 1 matter,''
  Phys.\ Lett.\ B {\bf 302}, 396 (1993)
  doi:10.1016/0370-2693(93)90416-F
  [hep-th/9301004].
 
\bibitem{Shenker:1990uf} 
  S.~H.~Shenker, 1990
  ``The Strength of nonperturbative effects in string theory,''
  In *Brezin, E. (ed.), Wadia, S.R. (ed.): The large N expansion in quantum field theory and statistical physics* 809-819


\bibitem{Belavin:2008kv}
A.~A.~Belavin and A.~B.~Zamolodchikov,
``On Correlation Numbers in 2D Minimal Gravity and Matrix Models,''
J. Phys. A \textbf{42}, 304004 (2009)
doi:10.1088/1751-8113/42/30/304004
[arXiv:0811.0450 [hep-th]].



\bibitem{Belavin:2010pj}
A.~Belavin and G.~Tarnopolsky,
``Two dimensional gravity in genus one in Matrix Models, Topological and Liouville approaches,''
JETP Lett. \textbf{92}, 257-267 (2010)
doi:10.1134/S0021364010160137
[arXiv:1006.2056 [hep-th]].



\bibitem{Belavin:2010bs}
A.~Belavin, M.~Bershtein and G.~Tarnopolsky,
``A remark on the three approaches to 2D Quantum gravity,''
JETP Lett. \textbf{93}, 47-51 (2011)
doi:10.1134/S0021364011020044
[arXiv:1010.2222 [hep-th]].

\bibitem{Gregori:2021tvs}
P.~Gregori and R.~Schiappa,
``From Minimal Strings towards Jackiw-Teitelboim Gravity: On their Resurgence, Resonance, and Black Holes,''
[arXiv:2108.11409 [hep-th]].


\bibitem{Ambjorn:2016lkl}
J.~Ambj\o{}rn, T.~Budd and Y.~Makeenko,
``Generalized multicritical one-matrix models,''
Nucl. Phys. B \textbf{913}, 357-380 (2016)
doi:10.1016/j.nuclphysb.2016.09.013
[arXiv:1604.04522 [hep-th]].


\bibitem{Tierz:2001kv}
M.~Tierz,
``The Stable random matrix ensembles,''
[arXiv:cond-mat/0106485 [cond-mat]].

 
\bibitem{Belavin:1984vu}
A.~A.~Belavin, A.~M.~Polyakov and A.~B.~Zamolodchikov,
``Infinite Conformal Symmetry in Two-Dimensional Quantum Field Theory,''
Nucl. Phys. B \textbf{241}, 333-380 (1984)
doi:10.1016/0550-3213(84)90052-X


  
 
\bibitem{Dotsenko:1985hi}
V.~S.~Dotsenko and V.~A.~Fateev,
``Operator Algebra of Two-Dimensional Conformal Theories with Central Charge C $\leq$ 1,''
Phys. Lett. B \textbf{154}, 291-295 (1985)
doi:10.1016/0370-2693(85)90366-1

\bibitem{Dotsenko:1984ad}
V.~S.~Dotsenko and V.~A.~Fateev,
``Four Point Correlation Functions and the Operator Algebra in the Two-Dimensional Conformal Invariant Theories with the Central Charge c \ensuremath{<} 1,''
Nucl. Phys. B \textbf{251}, 691-734 (1985)
doi:10.1016/S0550-3213(85)80004-3

\bibitem{Dotsenko:1986ca}
V.~S.~Dotsenko,
``LECTURES ON CONFORMAL FIELD THEORY,''
RIMS-559.


\bibitem{Cardy:1985yy}
J.~L.~Cardy,
``Conformal Invariance and the Yang-lee Edge Singularity in Two-dimensions,''
Phys. Rev. Lett. \textbf{54}, 1354-1356 (1985)
doi:10.1103/PhysRevLett.54.1354

\bibitem{Maloney}
S.~Collier, A.~Maloney, H.~Maxfield and I.~Tsiares,
``Universal dynamics of heavy operators in CFT$_{2}$,''
JHEP \textbf{07}, 074 (2020)
doi:10.1007/JHEP07(2020)074
[arXiv:1912.00222 [hep-th]].

\bibitem{Fitzpatrick:2014vua}
A.~L.~Fitzpatrick, J.~Kaplan and M.~T.~Walters,
``Universality of Long-Distance AdS Physics from the CFT Bootstrap,''
JHEP \textbf{08}, 145 (2014)
doi:10.1007/JHEP08(2014)145
[arXiv:1403.6829 [hep-th]].


\bibitem{Belavin:2003pu}
A.~A.~Belavin, V.~A.~Belavin, A.~V.~Litvinov, Y.~P.~Pugai and A.~B.~Zamolodchikov,
``On correlation functions in the perturbed minimal models M(2,2n+1),''
Nucl. Phys. B \textbf{676}, 587-614 (2004)
doi:10.1016/j.nuclphysb.2003.10.013
[arXiv:hep-th/0309137 [hep-th]].



\bibitem{wiki}
https://en.wikipedia.org/wiki/Euler-Maclaurin$\_$formula

\bibitem{Belin:2017nze}
A.~Belin, C.~A.~Keller and I.~G.~Zadeh,
``Genus two partition functions and R\'enyi entropies of large c conformal field theories,''
J. Phys. A \textbf{50}, no.43, 435401 (2017)
doi:10.1088/1751-8121/aa8a11
[arXiv:1704.08250 [hep-th]].



\bibitem{amoruso}
Nicola Amoruso,
``Renormalization group flows between non-unitary conformal models''. 


\bibitem{Friedan:1986ua}
D.~Friedan and S.~H.~Shenker,
``The Analytic Geometry of Two-Dimensional Conformal Field Theory,''
Nucl. Phys. B \textbf{281}, 509-545 (1987)
doi:10.1016/0550-3213(87)90418-4



\bibitem{Vafa:1988pw}
C.~Vafa,
``CONFORMAL ALGEBRA OF RIEMANN SURFACES,''
HUTP-88/A053.

\bibitem{Sonoda:1988mf}
H.~Sonoda,
``SEWING CONFORMAL FIELD THEORIES,''
Nucl. Phys. B \textbf{311}, 401-416 (1988)
doi:10.1016/0550-3213(88)90066-1

\bibitem{Sonoda:1988fq}
H.~Sonoda,
``SEWING CONFORMAL FIELD THEORIES. 2.,''
Nucl. Phys. B \textbf{311}, 417-432 (1988)
doi:10.1016/0550-3213(88)90067-3


\bibitem{Sen:1990bt}
A.~Sen,
``Some aspects of conformal field theories on the plane and higher genus Riemann surfaces,''
Pramana \textbf{35}, 205-286 (1990)
doi:10.1007/BF02846591


\bibitem{Zamolodchikov:2001dz}
A.~Zamolodchikov,
``Scaling Lee-Yang model on a sphere. 1. Partition function,''
JHEP \textbf{07}, 029 (2002)
doi:10.1088/1126-6708/2002/07/029
[arXiv:hep-th/0109078 [hep-th]].



\bibitem{Kapec:2020xaj}
D.~Kapec and R.~Mahajan,
``Comments on the quantum field theory of the Coulomb gas formalism,''
JHEP \textbf{04}, 136 (2021)
doi:10.1007/JHEP04(2021)136
[arXiv:2010.10428 [hep-th]].


\bibitem{David:1988hj} 
  F.~David,
  ``Conformal Field Theories Coupled to 2D Gravity in the Conformal Gauge,''
  Mod.\ Phys.\ Lett.\ A {\bf 3}, 1651 (1988).
  doi:10.1142/S0217732388001975
  
\bibitem{Distler:1988jt} 
  J.~Distler and H.~Kawai,
  ``Conformal Field Theory and 2D Quantum Gravity,''
  Nucl.\ Phys.\ B {\bf 321}, 509 (1989).
  doi:10.1016/0550-3213(89)90354-4
  
  
\bibitem{Goulian:1990qr}
M.~Goulian and M.~Li,
``Correlation functions in Liouville theory,''
Phys. Rev. Lett. \textbf{66}, 2051-2055 (1991)
doi:10.1103/PhysRevLett.66.2051

\bibitem{Ginsparg:1993is} 
  P.~H.~Ginsparg and G.~W.~Moore,
  ``Lectures on 2-D gravity and 2-D string theory,''
  Yale Univ. New Haven - YCTP-P23-92 (92,rec.Apr.93) 197 p. Los Alamos Nat. Lab. - LA-UR-92-3479 (92,rec.Apr.93) 197 p. e: LANL hep-th/9304011
  [hep-th/9304011].

  
\bibitem{DiFrancesco:1993cyw} 
  P.~Di Francesco, P.~H.~Ginsparg and J.~Zinn-Justin,
  ``2-D Gravity and random matrices,''
  Phys.\ Rept.\  {\bf 254}, 1 (1995)
  doi:10.1016/0370-1573(94)00084-G
  [hep-th/9306153].


\bibitem{Klebanov:1991qa}
I.~R.~Klebanov,
``String theory in two-dimensions,''
[arXiv:hep-th/9108019 [hep-th]].


\bibitem{Mahajan:2021nsd}
R.~Mahajan, D.~Stanford and C.~Yan,
``Sphere and disk partition functions in Liouville and in matrix integrals,''
[arXiv:2107.01172 [hep-th]].


\bibitem{Bershadsky:1990xb}
M.~Bershadsky and I.~R.~Klebanov,
``Genus one path integral in two-dimensional quantum gravity,''
Phys. Rev. Lett. \textbf{65}, 3088-3091 (1990)
doi:10.1103/PhysRevLett.65.3088

\bibitem{Bershadsky:1991zs}
M.~Bershadsky and I.~R.~Klebanov,
``Partition functions and physical states in two-dimensional quantum gravity and supergravity,''
Nucl. Phys. B \textbf{360}, 559-585 (1991)
doi:10.1016/0550-3213(91)90416-U
 

\bibitem{DHoker:1986eaw}
E.~D'Hoker and D.~H.~Phong,
``On Determinants of Laplacians on Riemann Surfaces,''
Commun. Math. Phys. \textbf{104}, 537 (1986)
doi:10.1007/BF01211063

\bibitem{Bonifacio:2020xoc}
J.~Bonifacio and K.~Hinterbichler,
``Bootstrap Bounds on Closed Einstein Manifolds,''
JHEP \textbf{10}, 069 (2020)
doi:10.1007/JHEP10(2020)069
[arXiv:2007.10337 [hep-th]].

\bibitem{Sarnak}
P. Sarnak, 
``Determinants of Laplacians''
Comm. Math. Phys. 110(1): 113-120 (1987)
https://doi.org/10.1007/BF01209019


\bibitem{Teschner:2003at}
J.~Teschner,
``From Liouville theory to the quantum geometry of Riemann surfaces,''
[arXiv:hep-th/0308031 [hep-th]].

\bibitem{Seiberg:1990eb}
N.~Seiberg,
``Notes on quantum Liouville theory and quantum gravity,''
Prog. Theor. Phys. Suppl. \textbf{102}, 319-349 (1990)
doi:10.1143/PTPS.102.319

\bibitem{Teschner:1995yf}
J.~Teschner,
``On the Liouville three point function,''
Phys. Lett. B \textbf{363}, 65-70 (1995)
doi:10.1016/0370-2693(95)01200-A
[arXiv:hep-th/9507109 [hep-th]].
 
\bibitem{Teschner:2001rv} 
  J.~Teschner,
  ``Liouville theory revisited,''
  Class.\ Quant.\ Grav.\  {\bf 18}, R153 (2001)
  doi:10.1088/0264-9381/18/23/201
  [hep-th/0104158].
 
\bibitem{Zamolodchikov:1995aa} 
  A.~B.~Zamolodchikov and A.~B.~Zamolodchikov,
  ``Structure constants and conformal bootstrap in Liouville field theory,''
  Nucl.\ Phys.\ B {\bf 477}, 577 (1996)
  doi:10.1016/0550-3213(96)00351-3
  [hep-th/9506136].
  
  \bibitem{Dorn:1994xn} 
  H.~Dorn and H.~J.~Otto,
  ``Two and three point functions in Liouville theory,''
  Nucl.\ Phys.\ B {\bf 429}, 375 (1994)
  doi:10.1016/0550-3213(94)00352-1
  [hep-th/9403141].
   
 
   
\bibitem{Harlow:2011ny}
D.~Harlow, J.~Maltz and E.~Witten,
``Analytic Continuation of Liouville Theory,''
JHEP \textbf{12}, 071 (2011)
doi:10.1007/JHEP12(2011)071
[arXiv:1108.4417 [hep-th]].

   \bibitem{PolchinskiBook}
 Polchinski, J. (1998). String Theory.  Vol. 1: An Introduction to the Bosonic String. (Cambridge Monographs on Mathematical Physics). 
 Cambridge: Cambridge University Press. doi:10.1017/CBO9780511816079

\bibitem{Douglas:1989dd}
M.~R.~Douglas,
``Strings in Less Than One-dimension and the Generalized $K^- D^- V$ Hierarchies,''
Phys. Lett. B \textbf{238}, 176 (1990)
doi:10.1016/0370-2693(90)91716-O

\bibitem{KPZ}
V.~G.~Knizhnik, A.~M.~Polyakov and A.~B.~Zamolodchikov,
``Fractal Structure of 2D Quantum Gravity,''
Mod. Phys. Lett. A \textbf{3}, 819 (1988)
doi:10.1142/S0217732388000982

\bibitem{Brezin:1989db}
E.~Brezin, M.~R.~Douglas, V.~Kazakov and S.~H.~Shenker,
``The Ising Model Coupled to 2-$D$ Gravity: A Nonperturbative Analysis,''
Phys. Lett. B \textbf{237}, 43-46 (1990)
doi:10.1016/0370-2693(90)90458-I

\bibitem{Holzhey:1994we} 
  C.~Holzhey, F.~Larsen and F.~Wilczek,
  ``Geometric and renormalized entropy in conformal field theory,''
  Nucl.\ Phys.\ B {\bf 424}, 443 (1994)
  doi:10.1016/0550-3213(94)90402-2
  [hep-th/9403108].
  
\bibitem{Calabrese:2004eu} 
  P.~Calabrese and J.~L.~Cardy,
  ``Entanglement entropy and quantum field theory,''
  J.\ Stat.\ Mech.\  {\bf 0406}, P06002 (2004)
  doi:10.1088/1742-5468/2004/06/P06002
  [hep-th/0405152].
  
\bibitem{Casini:2011kv} 
  H.~Casini, M.~Huerta and R.~C.~Myers,
  ``Towards a derivation of holographic entanglement entropy,''
  JHEP {\bf 1105}, 036 (2011)
  doi:10.1007/JHEP05(2011)036
  [arXiv:1102.0440 [hep-th]].
 


\bibitem{Shyam:2021ciy}
V.~Shyam,
``$T\bar{T}+\Lambda_2$ Deformed CFT on the Stretched dS$_3$ Horizon,''
[arXiv:2106.10227 [hep-th]].

\bibitem{Coleman:2021nor}
E.~Coleman, E.~A.~Mazenc, V.~Shyam, E.~Silverstein, R.~M.~Soni, G.~Torroba and S.~Yang,
``de Sitter Microstates from $T\bar T+\Lambda_2$ and the Hawking-Page Transition,''
[arXiv:2110.14670 [hep-th]].


\bibitem{Anninos:2017hhn}
D.~Anninos and D.~M.~Hofman,
``Infrared Realization of dS$_2$ in AdS$_2$,''
Class. Quant. Grav. \textbf{35}, no.8, 085003 (2018)
doi:10.1088/1361-6382/aab143
[arXiv:1703.04622 [hep-th]].


\bibitem{Anninos:2018svg}
D.~Anninos, D.~A.~Galante and D.~M.~Hofman,
``De Sitter horizons \& holographic liquids,''
JHEP \textbf{07}, 038 (2019)
doi:10.1007/JHEP07(2019)038
[arXiv:1811.08153 [hep-th]].

\bibitem{Anninos:2020cwo}
D.~Anninos and D.~A.~Galante,
``Constructing AdS$_{2}$ flow geometries,''
JHEP \textbf{02}, 045 (2021)
doi:10.1007/JHEP02(2021)045

\bibitem{Damian_Shira}
S.~Chapman, D.~A.~Galante and E.~D.~Kramer,
``Holographic Complexity and de Sitter Space,''
[arXiv:2110.05522 [hep-th]].

\bibitem{Susskind:2021esx}
L.~Susskind,
``Entanglement and Chaos in De Sitter Holography: An SYK Example,''
[arXiv:2109.14104 [hep-th]].

\bibitem{Shaghoulian:2021cef}
E.~Shaghoulian,
``The central dogma and cosmological horizons,''
[arXiv:2110.13210 [hep-th]].

\bibitem{Polchinski:1989fn}
J.~Polchinski,
``A Two-Dimensional Model for Quantum Gravity,''
Nucl. Phys. B \textbf{324}, 123-140 (1989)
doi:10.1016/0550-3213(89)90184-3

\bibitem{HH}
J.~B.~Hartle and S.~W.~Hawking,
Phys. Rev. D \textbf{28} (1983), 2960-2975
doi:10.1103/PhysRevD.28.2960



\bibitem{Anninos:2013nra}
D.~Anninos, T.~Anous, P.~de Lange and G.~Konstantinidis,
``Conformal quivers and melting molecules,''
JHEP \textbf{03}, 066 (2015)
doi:10.1007/JHEP03(2015)066
[arXiv:1310.7929 [hep-th]].

\bibitem{Kitaev:2017awl}
A.~Kitaev and S.~J.~Suh,
``The soft mode in the Sachdev-Ye-Kitaev model and its gravity dual,''
JHEP \textbf{05}, 183 (2018)
doi:10.1007/JHEP05(2018)183
[arXiv:1711.08467 [hep-th]].


\bibitem{Saad:2019lba}
P.~Saad, S.~H.~Shenker and D.~Stanford,
``JT gravity as a matrix integral,''
[arXiv:1903.11115 [hep-th]].

\bibitem{Coleman:1988tj}
S.~R.~Coleman,
``Why There Is Nothing Rather Than Something: A Theory of the Cosmological Constant,''
Nucl. Phys. B \textbf{310}, 643-668 (1988)
doi:10.1016/0550-3213(88)90097-1

\bibitem{Klebanov:1988eh}
I.~R.~Klebanov, L.~Susskind and T.~Banks,
``Wormholes and the Cosmological Constant,''
Nucl. Phys. B \textbf{317}, 665-692 (1989)
doi:10.1016/0550-3213(89)90538-5




%
%
%

%
%
  
  
\end{thebibliography}
\end{document}